\documentclass[12pt]{article}
\usepackage{a4wide,epsfig,psfrag,amsmath,amssymb,cite,scalefnt}
\usepackage[dvipsnames]{xcolor}
\usepackage{amsmath,comment,braket}
\usepackage{hyperref}
\usepackage{rotating}

\graphicspath{ {figs/} }

\newcommand{\arxhref}[1]{\href{https://arxiv.org/abs/#1}{#1}}  

\newcommand{\F}[1]{H^{(c)\,{#1}}}
\newcommand{\FS}[1]{\widetilde{H}_S^{(c)\,{#1}}}
\newcommand{\PP}[1]{H^{({c}p)\,{#1}}}
\newcommand{\PS}[1]{\widetilde{H}_S^{({c}p)\,{#1}}}
\newcommand{\PPP}[1]{H^{(p)\,{#1}}}
\newcommand{\PPS}[1]{\widetilde{H}_S^{(p)\,{#1}}}

\newcommand{\nn}{\nonumber\\}
\newcommand{\ov}{\overline}
\newcommand{\eq}[1]{Eq.~(\ref{#1})}
\newcommand{\eqsand}[2]{Eqs.~(\ref{#1}) and (\ref{#2})}
\newcommand{\eqsto}[2]{Eqs.~(\ref{#1}) to (\ref{#2})}
\newcommand{\gev}{\,\mbox{GeV}}

\newcommand{\Bbar}{\bar{B}}
\newcommand{\bbd}{\ensuremath{B_d\!-\!\Bbar{}_d\,}}
\newcommand{\bb}{\ensuremath{B\!-\!\Bbar\,}}
\newcommand{\bbs}{\ensuremath{B_s\!-\!\Bbar{}_s\,}}
\newcommand{\bbq}{\ensuremath{B_q\!-\!\Bbar{}_q\,}}
\newcommand{\bbm}{\bb\ mixing}
\newcommand{\bbms}{\bbs\ mixing}
\newcommand{\bbmd}{\bbd\ mixing}
\newcommand{\bbmq}{\bbq\ mixing}

\newcommand{\fig}[1]{Fig.~\ref{#1}}

\newcommand{\lqcd}{\Lambda_{\rm QCD}} 
\newcommand{\dm}{\ensuremath{\Delta M}}
\newcommand{\dg}{\ensuremath{\Delta \Gamma}}

\newcommand{\logOne} {L_1} 
\newcommand{\logTwo} {L_2} 

\allowdisplaybreaks

\begin{document}

\title{\vskip-3cm{\baselineskip14pt
    \begin{flushleft}
      \normalsize P3H-22-019, TTP22-012
    \end{flushleft}} \vskip1.5cm 
  The width difference in \bbm\ at order $\alpha_s$ and beyond
}
\author{
  Marvin Gerlach, Ulrich Nierste, Vladyslav Shtabovenko,
  \\
  and Matthias Steinhauser
  \\[1em]
  {\small\it Institut f{\"u}r Theoretische Teilchenphysik,
    Karlsruhe Institute of Technology (KIT)}\\
  {\small\it 76128 Karlsruhe, Germany}  
}
  
\date{}

\maketitle

\thispagestyle{empty}

\begin{abstract}
  We complete the calculation of the element $\Gamma_{12}^q$ of the
  decay matrix in $B_q-\bar{B}_q$ mixing, $q=d,s$, to order $\alpha_s$
  in the leading power of the Heavy Quark Expansion. To this end we
  compute one- and two-loop contributions involving two four-quark
  penguin operators. Furthermore, we present two-loop QCD corrections
  involving a chromomagnetic operator and either a current-current or
  four-quark penguin operator. Such contributions are of order
  $\alpha_s^2$, {\it i.e.}  next-to-next-to-leading-order. We also
  present one-loop and two-loop results involving two chromomagnetic
  operators which are formally of next-to-next-to-leading and
  next-to-next-to-next-to-leading-order, respectively.
  With our new corrections we obtain the Standard-Model prediction
    $\dg_s/\dm_s= (5.20\pm 0.69)\cdot 10^{-3}$ if  $\Gamma_{12}^s$
    is expressed in terms of the $\ov{\rm MS}$ b-quark mass,
    while we find
    $\dg_s/\dm_s= (4.70\pm 0.96)\cdot 10^{-3}$ instead for the use of the 
    pole mass.
\end{abstract}


\thispagestyle{empty}

\newpage


\section{Introduction}
The description of \bbmq, where $q=d$ or $s$, involves two hermitian
$2\times 2$ matrices, the mass matrix $M$ 
and the decay matrix $\Gamma$. Their off-diagonal elements
$M_{12}^q$ and  $\Gamma_{12}^q$ enter the observables related to
\bbmq, namely
\begin{eqnarray}
  \Delta M_q &=& M_H^q - M_L^q\,,\nonumber\\
  \Delta \Gamma_q &=& \Gamma_L^q - \Gamma_H^q\,,\label{eq:dg}\\
  \mbox{and~~} a_{\rm fs}^q &=& \mbox{Im} \frac{\Gamma_{12}^q}{M_{12}^q}\,. 
                 \label{eq::afsq}
\end{eqnarray}
Here $M_{L,H}$ and $\Gamma_{L,H}$ denote the masses and widths of the
two eigenstates found by diagonalizing $M-i\Gamma/2$.
The mass difference $\dm_q$ and the width difference $\dg_q$ are related
to $M_{12}^q$ and  $\Gamma_{12}^q$ as
\begin{eqnarray}
  \dm_q & \simeq & 2 |M_{12}^q|, \qquad\qquad
     \frac{\Delta\Gamma_q}{\Delta M_q} \;=\; - \mbox{Re}\frac{\Gamma_{12}^q}{M_{12}^q}\,.             
  \label{eq:dgdm}
\end{eqnarray}
In the Standard Model (SM) the phase between $-\Gamma_{12}^q$ and
$M_{12}^q$ is small, so that  the CP asymmetry in flavor-specific
decays, $ a_{\rm fs}^q$, is much smaller than
$\Delta\Gamma_q/\Delta M_q$ and further $\dg_q \simeq 2|\Gamma_{12}^q|$.

All results calculated in this paper equally apply to the
$B_s$ and $B_d$ systems. For definiteness, we quote all formulae for
the case of \bbms, the generalization to \bbmd\ is found by replacing
the elements $V_{qs}$, $q=u,c,t$, of the Cabibbo-Kobayashi-Maskawa (CKM) matrix 
by  $V_{qd}$. 

Currently, better theory predictions are needed for
the case of \bbms\ to be competitive with the precise experimental values
\begin{eqnarray}
  \dm^{\rm exp}_s &=&  (17.7656 \pm 0.0057)
                      \; \mbox{ps$^{-1}\qquad$\cite{LHCb:2021moh}}\,, \nn
  \dg^{\rm exp}_s  &=&  (0.082 \pm 0.005)\;
                       \mbox{ps}^{-1} \qquad
                       \hspace*{1.5em}\mbox{\cite{hfag}}  \label{eq:exp}\,,
\end{eqnarray}  
where the quoted number for $\dg^{\rm exp}_s$ is derived from data of LHCb
\cite{Aaij:2019vot}, CMS \cite{Sirunyan:2020vke}, ATLAS
\cite{ATLAS:2020lbz}, 
CDF \cite{CDF:2012nqr}, and D\O\ \cite{D0:2011ymu}.

$M_{12}^s$ probes virtual contributions of very heavy particles, while
{$\Gamma_{12}^s$} is mainly sensitive to new physics mediated by particles
with masses below the electroweak scale. Nevertheless, a better theory
prediction of $|\Gamma_{12}^s|$ also helps to quantify new physics in
$M_{12}^s$: Both $\dm_s$ and $\dg_s$ are proportional to $|V_{ts}|^2$,
where $V_{ts}$ is an element of the CKM
matrix.  $|V_{ts}|$ is calculated from (and is essentially identical to)
$|V_{cb}|$ extracted from measured $b\to c \ell\nu$, $\ell=e,\mu$,
branching ratios. The unfortunate discrepancy between the values for
$|V_{cb}|$ found from inclusive and exclusive decays inflicts an
uncertainty of order 15\% on the predicted $\dm_s$. Now $V_{ts}$ cancels
from $\dg_s/\dm_s$ in \eq{eq:dgdm}, so that the SM prediction of this
ratio is not affected by the $V_{cb}$ controversy.

In this paper we address {$\Gamma_{12}^s$} at leading order of the heavy-quark
expansion (``leading power''), which expresses {$\Gamma_{12}^s$} as a series in
powers of $\lqcd/m_b$.  At this order one encounters only two physical
$\Delta B=2$ operators, whose hadronic matrix elements have been calculated with
high precision with lattice QCD \cite{Dowdall:2019bea}. These matrix elements
are multiplied with Wilson coefficients which are calculated in perturbative
QCD. The insufficient accuracy of the Wilson coefficients dominates the uncertainty
of the SM prediction of $\dg_q$
\cite{Beneke:1998sy,Ciuchini:2003ww,Beneke:2003az,Lenz:2006hd,
  Asatrian:2017qaz,Asatrian:2020zxa,Gerlach:2021xtb}, which exceeds
the experimental error in \eq{eq:exp}. The perturbative calculation of power
corrections to $\Gamma_{12}^s$ has been carried out to order $\alpha_s^0$
\cite{Beneke:1996gn} and first lattice results for the associated hadronic matrix elements
are also available \cite{Davies:2019gnp}.

The $|\Delta B|=1$ Hamiltonian $\mathcal{H}_{\textrm{eff}}^{|\Delta B|=1}$
comprises current-current operators 
with large coefficients $C_{1,2}$ and the four-quark penguin operators whose
coefficients $C_{3-6}$ are small, with magnitudes well below 0.1, at the
scale $\mu_1={\cal O}(m_b)$ at which they enter $\Gamma_{12}^s$. At
order $\alpha_s^0$, $\Gamma_{12}^s$ is composed of one-loop
contributions proportional to $C_jC_k$ with $j,k\leq 6$.
$\mathcal{H}_{\textrm{eff}}^{|\Delta B|=1}$ further involves the
chromomagnetic penguin operator with coefficient $C_8\sim -0.16$, whose
leading contribution is of order $\alpha_s$ and enters $\Gamma_{12}^s$
as products $C_8C_k$ with $k\leq 6$ .  In
Refs.~\cite{Beneke:1998sy,Ciuchini:2003ww,Beneke:2003az,Lenz:2006hd,Asatrian:2017qaz,
  Asatrian:2020zxa} the small coefficients {$C_{3-6}$} have been formally
treated as ${\cal O}(\alpha_s)$. With this counting the one-loop terms with
{$C_{1,2} C_{3-6}$} contribute to next-to-leading order (NLO) and those
involving two factors of {$C_{3-6}$} are already part of the
next-to-next-to-leading order (NNLO). First steps towards NNLO accuracy have
been done in Refs.~\cite{Asatrian:2017qaz, Asatrian:2020zxa} by calculating
contribution proportional to the number of active quark flavors, i.e.\ loop
diagrams with a closed fermion line.

As in Ref.~\cite{Gerlach:2021xtb} we use the conventional notion of ``NLO'' and
``NNLO'' in this paper and treat {$C_{3-6}$} on the same footing as $C_{1,2}$. With this
counting the NLO prediction of $\Gamma_{12}^s$ requires the calculation of the
yet unknown two-loop contributions with one or two four-quark penguin operators.
In this paper present several two-loop calculations, namely: 
\begin{itemize}
\item penguin contributions proportional to the product of two $C_{3-6}$
  coefficients. This contribution completes the prediction of
  $\Gamma_{12}^s$ to order $\alpha_s$, which is NLO in the
  above-mentioned conventional power counting. The corresponding
  one-loop corrections have been computed in
  Ref.~\cite{Beneke:1996gn}. Two-loop contributions with one
  current-current and one four-quark penguin operator have been
  calculated in Ref.~\cite{Gerlach:2021xtb}.

\item the contribution proportional to the product of $C_8$ and one of
  $C_{1-6}$. The calculated one-loop and two-loop terms contribute
  to NLO and NNLO, respectively.  The piece of the one-loop
  correction proportional to the number $N_f$ of active quark flavors
  (stemming from diagrams with closed quark loops) has been computed
  in Ref.~\cite{Asatrian:2020zxa}.
    
\item the contribution proportional to $C_8^2$.
  Here the one-loop contribution is already of NNLO and not yet
  available in the literature, except for the $\alpha_s^2N_f$ part
  \cite{Asatrian:2020zxa}. We further provide results for the
  two-loop term which is N$^3$LO.
\end{itemize}

As in Ref.~\cite{Gerlach:2021xtb} we use the CMM basis
\cite{Chetyrkin:1997gb} for the $|\Delta B|=1$ operators and calculate
the two-loop QCD corrections as an expansion in
\begin{eqnarray}
  z &=& \frac{m_c^2}{m_b^2}          \label{eq:defz}
\end{eqnarray}
up to linear order.

The paper is organized as follows: In the next Section we briefly
discuss the operator bases of the $|\Delta B|=1$ and $|\Delta B|=2$
theories. Afterwards, in Section~\ref{sec::calc} we provide some details
of our calculation and in particular describe the matching procedure for
the case of dimensionally regularized infra-red singularities.  Analytic
result for all new matching coefficients are listed in
Section~\ref{sec::ana} and we present our numerical result for $\dg_s$
in Section~\ref{sec::num}.  Section~\ref{sec::concl} contains our
conclusions. In the Appendix we provide results for the renormalization
constants relevant for the operator mixing in the $|\Delta B|=2$ theory.


\section{Operator bases\label{sec:pr}}

The framework of our calculation is identical to the one used in
Ref.~\cite{Gerlach:2021xtb} and thus in the following we repeat 
only the essential formulae needed to compute the width difference.
The new contributions considered in this paper require an extension of the 
$|\Delta B|=2$ operator basis which is discussed in more detail.

For the effective $|\Delta B|=1$ theory we use the weak Hamiltonian
\begin{eqnarray}
  \mathcal{H}_{\textrm{eff}}^{|\Delta B|=1}
  &=&   \frac{4G_F}{\sqrt{2}}  \left[
      -\, \lambda^s_t \Big( \sum_{i=1}^6 C_i Q_i + C_8 Q_8 \Big) 
      - \lambda^s_u \sum_{i=1}^2 C_i (Q_i - Q_i^u) \right. \nn
  && \phantom{\frac{4G_F}{\sqrt{2}} \Big[}
      \left.
      +\, V_{us}^\ast V_{cb} \, \sum_{i=1}^2 C_i Q_i^{cu} 
      + V_{cs}^\ast V_{ub} \, \sum_{i=1}^2 C_i Q_i^{uc} 
      \right]
      + \mbox{h.c.}\,,
      \label{eq::HamDB1}
\end{eqnarray}
where explicit expressions for the (physical and evanescent) operators can be
found in Ref.~\cite{Chetyrkin:1997gb}.  $Q_1, Q_1^{{(u,cu,uc)}}, Q_2$ and $Q_2^{{(u,cu,uc)}}$
are current-current and $Q_3,\ldots,Q_6$ are four-quark penguin operators. $Q_8$
is the chromomagnetic penguin operator.  In Eq.~(\ref{eq::HamDB1}) we have introduced
the quantities $\lambda^s_a = V_{as}^\ast V_{ab}$, $a=u,c,t,$ which contain
the CKM matrix elements. Furthermore, we have used $\lambda^s_t=-\lambda^s_c-\lambda^s_u$
and $G_F$ is the Fermi constant. Our two-loop calculations involve
one-loop diagrams with counterterms to the physical operators in
\eq{eq::HamDB1} and these counterterms comprise both physical and
evanescent operators. 
 
As mentioned in the Introduction, we specify our discussion to $b\to s$ decays relevant for
\bbms. The corresponding expressions for \bbmd\ are obtained by replacing
$V_{as}$ with $V_{ad}$. Using the optical theorem we can relate
$\Gamma_{12}^s$ to the $\bar B_s \to B_s$ forward scattering amplitude: 
\begin{eqnarray}
  \Gamma_{12}^s &=& \frac{1}{2 M_{B_s}} \,\mbox{Abs}\langle B_s|i\int{\rm d}^4 x \,\, T\,\,
                    {\cal H}_{\rm eff}^{\Delta B=1}(x)
                    {\cal H}_{\rm eff}^{\Delta B=1}(0)
                    |\bar{B}_s\rangle\,, \label{eq:ot}
\end{eqnarray}
where ``Abs'' stands for the absorptive part and $T$ is the time ordering
operator. Note that $\Gamma_{12}^s$ encodes the information of the inclusive
decay rate into final states common to $B_s$ and $\bar B_s$.
Following Ref.~\cite{Beneke:1998sy} we decompose $\Gamma_{12}^s$ as
\begin{eqnarray}
  \Gamma_{12}^s &=& - (\lambda_c^s)^2\Gamma^{cc}_{12} 
                  - 2\lambda_c^s\lambda_u^s \Gamma_{12}^{uc} 
                  - (\lambda_u^s)^2\Gamma^{uu}_{12} 
                  \,.
                    \label{eq::Gam12}
\end{eqnarray}

Let us now discuss the effective $|\Delta B|=2$ theory.
To leading power in $1/m_b$ it is convenient to introduce
the following four operators
\begin{eqnarray}
  Q &=& \bar{s}_i \gamma^\mu \,(1-\gamma^5)\, b_i \; \bar{s}_j \gamma_\mu
        \,(1-\gamma^5)\, b_j\,, \nonumber\\
  \widetilde{Q} &=& \bar{s}_i \gamma^\mu \,(1-\gamma^5)\, b_j \; \bar{s}_j \gamma_\mu
                 \,(1-\gamma^5)\, b_i\,,\nonumber\\
  \widetilde{Q}_S &=& \bar{s}_i \,(1+\gamma^5)\, b_j\; \bar{s}_j \,(1+\gamma^5)\,
                  b_i\, 
                      \nonumber\\
  Q_S &=& \bar{s}_i \,(1+\gamma^5)\, b_i \;\bar{s}_j \,(1+\gamma^5)\,
         b_j\,,
          \label{eq::opDB2}
\end{eqnarray}
where $i,j$ are color indices.  In four space-time dimensions there are
only two independent operators which we choose as $Q$ and $\widetilde{Q}_S$
since we have (for $D=4$) $Q=\widetilde{Q}$ and
\begin{eqnarray}
  Q_S &=& -{\alpha_1} \widetilde Q_S - \frac{1}{2} {\alpha_2} Q
          + R_0\,, \label{eq:defr0}
         \label{eq::R0}
\end{eqnarray}  
where $R_0$ describes $1/m_b$-suppressed contributions to $\Gamma_{12}^s$
\cite{Beneke:1996gn}. $\alpha_{1,2}$ are QCD correction factors
which ensure that the $\ov{\rm MS}$ renormalized matrix element
$\langle R_0\rangle$ has the desired power suppression
\cite{Beneke:1998sy,Lenz:2006hd}. 

Using the Heavy Quark Expansion (HQE) it is thus possible to write
$\Gamma_{12}^{ab}$ in \eq{eq::Gam12} as
\begin{eqnarray}
  \Gamma_{12}^{ab} 
  &=& \frac{G_F^2m_b^2}{24\pi M_{B_s}} \left[ 
      H^{ab}(z)   \langle B_s|Q|\bar{B}_s \rangle
      + \widetilde{H}^{ab}_S(z)  \langle B_s|\widetilde{Q}_S|\bar{B}_s \rangle
      \right]
      + \ldots \,
      \label{eq::Gam^ab}
\end{eqnarray}
with $z=(m_c^{\rm pole}/m_b^{\rm pole})^2$ and
the ellipses denoting higher-order terms in $\Lambda_{\rm QCD}/m_b$.
Here $z$ is defined in terms of pole quark masses.
Later we will trade $z$ for the ratio of $\ov{\rm MS}$ masses
which leads to a better behavior of the perturbative series.
$H^{ab}$ and $\widetilde{H}^{ab}_S$ are ultra-violet and infra-red finite
matching coefficients which we decompose as follows
\begin{eqnarray}
  H{}^{ab}(z) &=& \F{ab} (z)+ \PP{ab}(z)  +\PPP{ab}(z) 
                  \,, \nonumber\\
  \widetilde{H}_S^{ab}(z) 
              &=& \FS{ab}(z)+ \PS{ab}(z) +\PPS{ab}(z)
                  \,, 
                  \label{eq::G}
\end{eqnarray}
where the superscript ``(c)'' denotes the contributions with two
current-current operators $Q_{1,2}$ or $Q_{1,2}^{(u,cu,uc)}$, ``(cp)''
refers to those with one operator $Q_{1,2}$ or $Q_{1,2}^{(u,cu,uc)}$ and
one (four-quark or chromomagnetic) penguin operator $Q_{3-6,8}$ and ``(p)''
labels the terms involving two penguin operators.
In this paper we present new contributions to
$\PPP{ab}$ and $\PPS{ab}(z)$
up to two-loop order.

At intermediate steps (i.e. in $D=4-2\epsilon$ dimensions) of our calculation it is
convenient to use all four operators of Eq.~(\ref{eq::opDB2}) together with
evanescent operators with two or three Dirac matrix structures given
by~\cite{Beneke:1998sy,Gorbahn:2009pp}
\begin{eqnarray}
  E_1^{(1)} &=& \widetilde{Q} - Q\,, \nonumber\\
  E_2^{(1)} &=& \bar{s}_i \gamma^\mu \gamma^\nu \gamma^\rho \,{(1-\gamma_5)}\,
                {b}_j\;
          \bar{s}_j \gamma_\mu \gamma_\nu \gamma_\rho \,{(1-\gamma_5)}\, {b}_i - (16 -4 \epsilon) \widetilde{Q}\,, \nonumber\\
  E_3^{(1)} &=& \bar{s}_i \gamma^\mu \gamma^\nu \gamma^\rho \,{(1-\gamma_5)}\,
                {b}_i  \;\bar{s}_j \gamma_\mu \gamma_\nu \gamma_\rho \,{(1-\gamma_5)}\, {b}_j - (16 -4 \epsilon) Q\,, \nonumber\\
  E_4^{(1)} &=& \bar{s}_i \gamma^\mu \gamma^\nu \,{(1+\gamma_5)}\, {b}_j
                \;\bar{s}_j \gamma_\nu \gamma_\mu \,{(1+\gamma_5)}\, {b}_i  + (8 - 8 \epsilon) Q_s\,, \nonumber\\
  E_5^{(1)} &=& \bar{s}_i \gamma^\mu \gamma^\nu \,{(1+\gamma_5)}\, {b}_i
                \;\bar{s}_j \gamma_\nu \gamma_\mu \,{(1+\gamma_5)}\, {b}_j  + (8 - 8 \epsilon) \widetilde{Q}_s\,.
              \label{eq:E1}
\end{eqnarray}
The ${\cal O}(\epsilon)$ parts in Eq.~(\ref{eq:E1}) are chosen such that the
Fierz symmetry of the renormalized $|\Delta B|=2$ amplitudes extends to $D$
dimensions \cite{Herrlich:1994kh} and ${\cal O}(\epsilon^2)$ terms,
which are important for a three-loop (NNLO)
calculation, have been omitted. Furthermore, we remark that the five evanescent operators in
Eq.~(\ref{eq:E1}) are needed in order to determine the renormalization
constants responsible for the operator mixing in the $|\Delta B|=2$ theory, see
Appendix~\ref{app::Z}.

\begin{table}[t]
  \begin{center}
    \begin{tabular}{rcl|c}
      \hline 
      \multicolumn{3}{c|}{Contribution}
                   &   Maximal number of $\gamma$ matrices needed\\
                   &&& for the two-loop calculation     \\
      \hline 
      $Q_{1,2}$ &$\times$& $Q_{1,2}$ & $5 \times 5$ \\
      $Q_{1,2}$ &$\times$& $Q_{3-6}$ & $5 \times 5$ \\
      $Q_{3-6}$ &$\times$& $Q_{3-6}$ & $9 \times 9$ \\
      $Q_{1,2}$ &$\times$& $Q_{8}$   & $3 \times 3$ \\
      $Q_{3,6}$ &$\times$& $Q_{8}$   & $7 \times 7$ \\
      $Q_{8}$   &$\times$& $Q_{8}$   & $5 \times 5$ \\
      \hline 
    \end{tabular} 
    \caption{\label{tab::gamma_structures}
      Maximal number of $\gamma$ matrices which appear
      in the calculation of two-loop corrections to the various contributions
      involving current-current and penguin operators.}
  \end{center}
  ~\\[-4mm] \hrule
\end{table}

In our calculation we encounter further evanescent $|\Delta B|=2$
operators, since in intermediate steps Dirac structures with up to nine
different $\gamma$ matrices can appear. In
Tab.~\ref{tab::gamma_structures} we list the maximal number of $\gamma$
matrices for each pair of $|\Delta B|=1$ operators. It is easily obtained by
inspecting the corresponding one-loop diagrams with one physical and one
evanescent operator from Eqs.~(\ref{eq::opDB2}) and~(\ref{eq:E1}),
respectively, or two-loop diagrams with two physical operators. We
define the additional evanescent operators as
\begin{eqnarray}
  E_1^{(2)} &=& \bar{s}_i \gamma^{\mu_1} \ldots \gamma^{\mu_5} \,{(1-\gamma_5)}\, {b}_j 
                \;\bar{s}_j \gamma_{\mu_1} \ldots \gamma_{\mu_5} \,{(1-\gamma_5)}\, {b}_i  
                - (256 + {e}_1^{(2)} \epsilon ) \tilde{Q}\,, \nonumber\\ 
  E_2^{(2)} &=& \bar{s}_i \gamma^{\mu_1} \ldots \gamma^{\mu_5} \,{(1-\gamma_5)}\, {b}_i 
                \;\bar{s}_j \gamma_{\mu_1} \ldots \gamma_{\mu_5} \,{(1-\gamma_5)}\, {b}_j  
                - (256 + {e}_2^{(2)} \epsilon ) Q\,, \nonumber\\ 	 
  E_3^{(2)} &=& \bar{s}_i \gamma^{\mu_1} \ldots \gamma^{\mu_4} \,{(1+\gamma_5)}\, {b}_i  
                \;\bar{s}_j \gamma_{\mu_1} \ldots \gamma_{\mu_4} \,{(1+\gamma_5)}\, {b}_j  
                - (128 + {e}_{3,1}^{(2)} \epsilon ) \tilde{Q}_S 
\nonumber\\&&\mbox{}
- (128 + {e}_{3,2}^{(2)} \epsilon
                ) Q_S\,, \nonumber\\ 
  E_4^{(2)} &=& \bar{s}_i \gamma^{\mu_1} \ldots \gamma^{\mu_4} \,{(1+\gamma_5)}\, {b}_j   
                \;\bar{s}_j \gamma_{\mu_1} \ldots \gamma_{\mu_4} \,{(1+\gamma_5)}\, {b}_i   
                - (128 + {e}_{4,1}^{(2)} \epsilon ) \tilde{Q}_S
\nonumber\\&&\mbox{}
 - (128 + {e}_{4,2}^{(2)} \epsilon
                ) Q_S\,,
                \nonumber\\
  E_1^{(3)} &=& \bar{s}_i \gamma^{\mu_1} \ldots \gamma^{\mu_7} \,{(1-\gamma_5)}\,
                {b}_j \;\bar{s}_j \gamma_{\mu_1} \ldots \gamma_{\mu_7}
                \,{(1-\gamma_5)}\, {b}_i - (4096 + {e}_1^{(3)} \epsilon ) \tilde{Q}, \nonumber\\ 
  E_2^{(3)} &=& \bar{s}_i \gamma^{\mu_1} \ldots \gamma^{\mu_7} \,{(1-\gamma_5)}\,
                {b}_i \;\bar{s}_j \gamma_{\mu_1} \ldots \gamma_{\mu_7}
                \,{(1-\gamma_5)}\, {b}_j - (4096 + {e}_2^{(3)} \epsilon ) Q, \nonumber\\
  E_3^{(3)} &=& \bar{s}_i \gamma^{\mu_1} \ldots \gamma^{\mu_6}
		\,{(1+\gamma_5)}\, {b}_i \;\bar{s}_j \gamma_{\mu_1} \ldots \gamma_{\mu_6}
		\,{(1+\gamma_5)}\, {b}_j  - (2048 + {e}_{3,1}^{(3)} \epsilon )
                \tilde{Q}_S 
\nonumber\\&&\mbox{}
-
                (2048 + {e}_{3,2}^{(3)} \epsilon ) Q_S, \nonumber\\ 
  E_4^{(3)} &=& \bar{s}_i \gamma^{\mu_1} \ldots \gamma^{\mu_6}
		\,{(1+\gamma_5)}\, {b}_j \;\bar{s}_j \gamma_{\mu_1} \ldots \gamma_{\mu_6}
		\,{(1+\gamma_5)}\, {b}_i   - (2048 + {e}_{4,1}^{(3)} \epsilon
                ) \tilde{Q}_S 
\nonumber\\&&\mbox{}
-
                (2048 + {e}_{4,2}^{(3)} \epsilon ) Q_S\,, 
  \nonumber\\
  E_1^{(4)} &=& \bar{s}_i \gamma^{\mu_1} \ldots \gamma^{\mu_9} \,{(1-\gamma_5)}\,
                {b}_j \;\bar{s}_j \gamma_{\mu_1} \ldots \gamma_{\mu_9}
                \,{(1-\gamma_5)}\, {b}_i - (65536 + {e}_1^{(4)} \epsilon ) \tilde{Q}, \nonumber\\  
  E_2^{(4)} &=& \bar{s}_i \gamma^{\mu_1} \ldots \gamma^{\mu_9} \,{(1-\gamma_5)}\,
                {b}_i \;\bar{s}_j \gamma_{\mu_1} \ldots \gamma_{\mu_9}
                \,{(1-\gamma_5)}\, {b}_j - (65536 + {e}_2^{(4)} \epsilon ) Q, \nonumber\\
  E_3^{(4)} &=& \bar{s}_i \gamma^{\mu_1} \ldots \gamma^{\mu_8}
		\,{(1+\gamma_5)}\, {b}_i \;\bar{s}_j \gamma_{\mu_1} \ldots \gamma_{\mu_8}
                \,{(1+\gamma_5)}\, {b}_j  - (32768  + {e}_{3,1}^{(4)} \epsilon ) \tilde{Q}_S \nonumber \\
                &&\mbox{} -(32768 + {e}_{3,2}^{(4)} \epsilon ) Q_S, \nonumber\\  
  E_4^{(4)} &=& \bar{s}_i \gamma^{\mu_1} \ldots \gamma^{\mu_8}
		\,{(1+\gamma_5)}\, {b}_j \;\bar{s}_j \gamma_{\mu_1} \ldots \gamma_{\mu_8}
		\,{(1+\gamma_5)}\, {b}_i   - (32768  + {e}_{4,1}^{(4)} \epsilon ) \tilde{Q}_S \nonumber \\
				&&\mbox{}- (32768 + {e}_{4,2}^{(4)} \epsilon ) Q_S\,,
                \label{eq::evOp_add}
\end{eqnarray}
where for our calculation the values of ${e}_j^{(k)}$,${e}_{j,l}^{(k)}$,
which parametrize the ${\cal O}(\epsilon)$ terms in the definition of
the evanescent operators are irrelevant since the operators in
Eq.~(\ref{eq::evOp_add}) do not appear in one-loop counterterm
contributions.  These numbers, however, become important at NNLO to
fully specify the renormalization scheme at this order.  


\section{\label{sec::calc}Calculation and Matching}

The setup which we use for our calculation has already been described in
Ref.~\cite{Gerlach:2021xtb}. For convenience of the reader we repeat the
essential steps and stress the differences in the following.

\begin{figure}[t]
  \begin{center}
    \includegraphics[width=\textwidth]{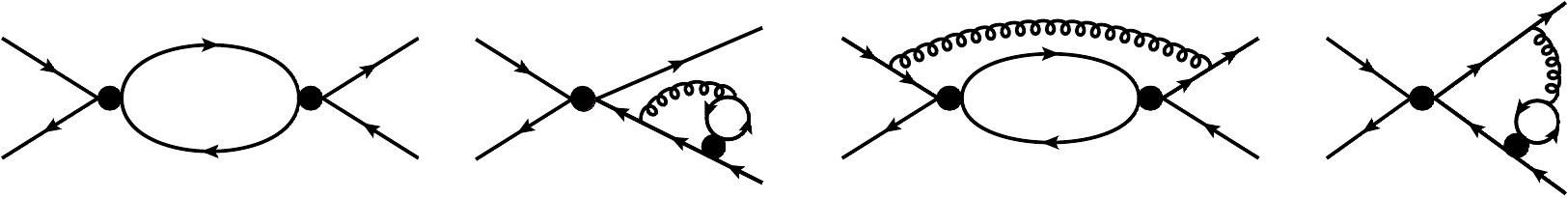} \\
    \includegraphics[width=\textwidth]{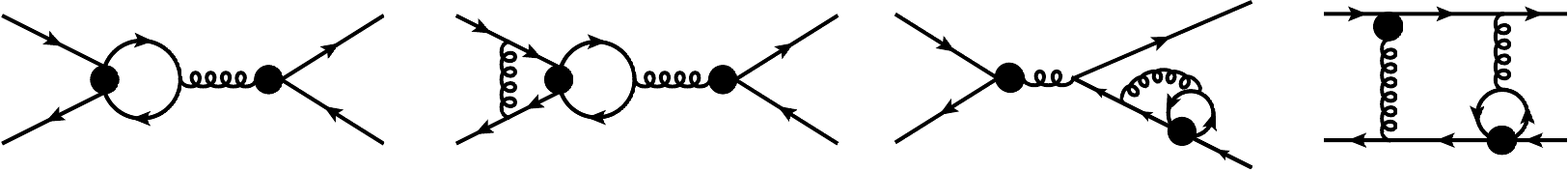} \\
    \includegraphics[width=\textwidth]{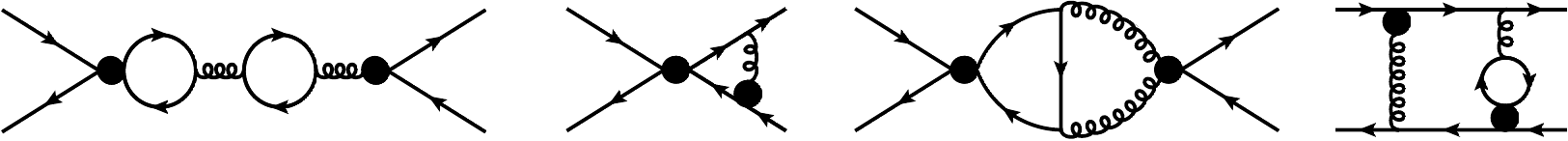} \\
    \includegraphics[width=\textwidth]{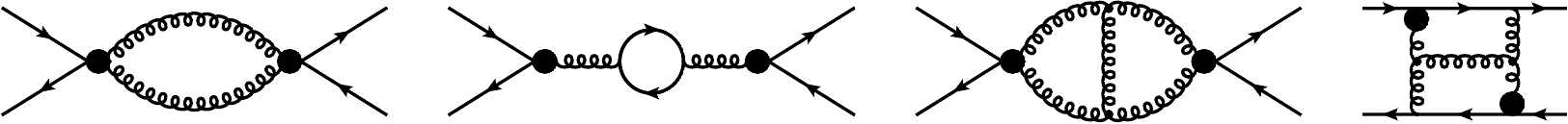}
  \end{center}
  \caption{\label{fig::digs}Sample Feynman diagrams contributing to
    \eq{eq:ot} to the orders considered in this paper. From top to
    bottom they contribute to the $Q_{3-6}\times Q_{3-6}$,
    $Q_{1,2}\times Q_{8}$, $Q_{3,6}\times Q_{8}$, and
    $Q_{8}\times Q_{8}$ pieces of
    ${\cal H}_{\rm eff}^{\Delta B=1}(x) {\cal H}_{\rm eff}^{\Delta
      B=1}(0)$ in \eq{eq:ot}, with the blobs denoting the corresponding
    current-current or penguin operators.  }
~\\[-4mm] \hrule
\end{figure}

Figure~\ref{fig::digs} shows typical one- and two-loop Feynman diagrams for
the new contributions considered in this paper. The displayed diagrams
correspond to the $|\Delta B|=1$ side of the matching equation. In addition,
one needs the one-loop diagrams with a gluon dressing the $\Delta B=2$
operators $Q$ and $\tilde{Q}_S$ to determine the desired Wilson coefficients
$H^{ab}$ and $\widetilde{H}^{ab}_S$ in \eqsand{eq::Gam^ab}{eq::G}.  We perform
the calculation for a generic QCD gauge parameter which drops out in the final
result for each matching coefficient and thereby provides a non-trivial check
of our calculation.  The counterterms to the $\Delta B = 1$ operators and the
gauge coupling $g_s$ in the Feynman diagrams exemplified in \fig{fig::digs}
are all evaluated at the renormalization scale $\mu_1$. Conversely, operators
and couplings on the $|\Delta B|=2$ side are evaluated at the scale
$\mu_2$. The unphysical $\mu_1$ dependence of $H^{ab}(z)$ and
$\widetilde{H}_S^{ab}(z)$ diminishes order-by-order in perturbation theory and
can be used to assess the accuracy of the calculated result. The $\mu_2$
dependence of $H^{ab}(z)$ and $\widetilde{H}_S^{ab}(z)$, however, cancels with
the $\mu_2$ dependence of the hadronic matrix element, which enters the
lattice-continuum matching. For calculational convenience we first choose
$\mu_1=\mu_2$ and implement the separation $\mu_1\not=\mu_2$ with the help of
renormalization group techniques.

We pursue two different approaches to treat the four-quark
amplitudes. The first one is based on tensor integrals combined with
various manipulations of the Dirac structures and relies on {\tt
  FeynCalc}~\cite{Mertig:1990an,Shtabovenko:2016sxi,Shtabovenko:2020gxv}
and {\tt Fermat}~\cite{fermat}. The so-obtained formulae are then exported to \texttt{FORM}~\cite{Kuipers:2012rf}.

 For the contribution
$Q_{3-6}\times Q_{3-6}$ routines are needed which can handle tensor
integrals up to rank~6.  The second approach is based on
projectors (see Appendix of Ref.~\cite{Gerlach:2021xtb}) which allows
taking traces. Thus, one only has to deal with scalar expressions. However,
one needs to calculate products of two traces with up to 18
$\gamma$ matrices\footnote{Up to nine $\gamma$ matrices are present in
  the (two-loop) amplitude (see Tab~\ref{tab::gamma_structures}) and
  nine $\gamma$ matrices come from the projector.} in each trace.  We
find that both approaches lead to the same expressions for the amplitude
with two $\Delta B=1$ operators once the latter is expressed in terms of
tree-level $\Delta B=2$ matrix elements.

For the reduction of the $\Delta B=1$ amplitude we use {\tt
  FIRE}~\cite{Smirnov:2019qkx} with symmetries from {\tt
  LiteRed}~\cite{Lee:2012cn,Lee:2013mka} and obtain four two-loop master
integrals.  Their evaluation as an expansion in $\epsilon$ is straightforward.

The amplitudes in the $|\Delta B|=1$ and $|\Delta B|=2$ theories contain both
ultra-violet and infra-red singularities. The former are cured with 
parameter, quark field, and operator renormalization. 
We use the one-loop counterterms for
$\alpha_s$ in the $\overline{\rm MS}$ scheme and renormalize the charm quark
in the one-loop expression in the on-shell (or pole) scheme. The
renormalization of the bottom quark, which we also renormalize on-shell, is
only needed for the contributions involving $Q_8$. We also perform
the renormalization of the external quark fields in the $\overline{\rm MS}$
scheme. The counterterms needed for the renormalization
of the  $\Delta B=1$ operator mixing can be taken from the
literature~\cite{Gambino:2003zm}. The  renormalization constants
of the $\Delta B=2$ part are given in Appendix~\ref{app::Z}.

In order to regulate the infra-red singularities two possibilities come
to mind: One can either introduce a (small) gluon mass, $m_g$, or
instead use dimensional regularization.

The choice $m_g\not=0$ is conceptually simpler and has the advantage that
after renormalization the $\Delta B =1$ and $\Delta B =2$ amplitudes are
separately finite and one can take the limit $D\to4$, which eliminates all
evanescent operators before matching the two theories.  Furthermore, it is
possible to use four-dimensional relations in order to arrive at a minimal
operator basis.  However, a finite gluon mass breaks gauge invariance and
thus, in general, additional counterterms have to be introduced for its
restoration.  In our application the two-loop $\Delta B =1$ amplitudes
with four-quark operators do not involve three-gluon vertices, and thus it
is safe to regulate the infra-red divergences with $m_g\not=0$.  However, at
three-loop level this is not the case. Furthermore, the two-loop corrections with
two $Q_8$ operators also contain infra-red divergences in the non-abelian
part.

Regulating the infra-red divergences dimensionally using the same
regulator $\epsilon$ as for the ultra-violet divergences has the
advantage that the loop integrals are simpler. However, the matching has
to be performed with divergent quantities in $D\not=4$ dimensions. As a
consequence lower-order corrections have to be computed to higher order
in $\epsilon$, meaning that also the evanescent operators have to
be taken into account.

In our calculation we proceeded as follows: We have computed the
contributions $Q_{1-6}\times Q_{1-6}$ and $Q_{1-2}\times Q_{8}$ both for
$m_g\not=0$ and $m_g=0$ and have obtained identical results for the 
matching coefficients, which provides sufficient confidence that
the conceptually more involved approach where the infra-red divergences
are regularized dimensionally is understood. Thus, the calculation
of the $Q_{3-6}\times Q_{8}$ and $Q_{8}\times Q_{8}$ have only
been performed for $m_g=0$.

In the following we provide some details to the matching procedure.  In this
context we also refer to Ref.~\cite{Ciuchini:2001vx} where
the contribution $Q_{1,2} \times Q_{1,2}$ is discussed.
We introduce the $|\Delta B|=1$ and $|\Delta B|=2$ amplitude in a schematic way as
\begin{eqnarray}
  {\cal A}^{\Delta B=1} &=&   A_{Q} \langle Q   \rangle^0 
                            + A_{E} \langle {E} \rangle^0
                            \,,\nonumber\\
  {\cal A}^{\Delta B=2} &=&   H_Q B_{QQ} \langle Q   \rangle^0 
                            + H_E B_{EQ} \langle Q   \rangle^0
                            + H_Q B_{QE} \langle {E} \rangle^0
                            + H_E B_{EE} \langle {E} \rangle^0
                            \,, \label{eq:schematic}
\end{eqnarray}
where the $H_{X}$, $A_{X}$ and $B_{XY}$ have an expansion both in $\alpha_s$
and $\epsilon$ with $B_{QQ}=B_{EE}=1$ and $B_{EQ}=B_{QE}=0$ at LO.  
$\langle \cdot \rangle^0$ denote tree-level
matrix elements. Starting
from two-loop order\footnote{The counting of loop orders always refers to the
  $|\Delta B|=1$ side of the matching equation.}  $A_{X}$ and $B_{XY}$ contain
infrared $1/\epsilon$ poles. The presence of these poles force us to
  calculate the LO coefficients $H_{X}$ to order $\epsilon$ in order to
  obtain the correct finite $\epsilon^0$ piece on the right-hand side of
  \eq{eq:schematic}.
Thus, the desired finite matching coefficients  $H_{X}$ have the 
following expansion in $\alpha_s$ and $\epsilon$:
\begin{eqnarray}
  H_Q &=& \sum_{i,j\ge0} H_Q^{(i,j)} \epsilon^j \left(\frac{\alpha_s}{4\pi}\right)^i
          \,,
\end{eqnarray}
and analogously for $H_E$. In general, several physical (``$Q$'') and evanescent
(``$E$'') operators are present; for simplicity we condense the notation to
only one operator for in each case.

We start the matching at LO,
which corresponds to a one-loop calculation of $A_{Q}$ and $A_{E}$.  For
$B_{XY}$ we use the tree-level expressions.  Both ${\cal A}^{\Delta B=1}$ and
${\cal A}^{\Delta B=2}$ are finite and from the comparison of both amplitudes
we obtain results for $H_Q^{(0,0)}$, $H_Q^{(0,1)}$, $H_E^{(0,0)}$ and
$H_E^{(0,1)}$.

At NLO we observe that after using the result for $H_Q^{(0,0)}$ and
$H_E^{(0,0)}$ the difference
${\cal A}^{\Delta B=1}-{\cal A}^{\Delta B=2}$ is finite, which
constitutes an important consistency check.  In a next step we
concentrate on the part of ${\cal A}^{\Delta B=1}-{\cal A}^{\Delta B=2}$
proportional to $\langle Q \rangle^0$, which contains $H_Q^{(1,0)}$ as
the desired finite coefficient.  $H_Q^{(1,0)}$ can thus be determined by
requiring the $\langle Q \rangle^0$ part of
${\cal A}^{\Delta B=1}-{\cal A}^{\Delta B=2}$ to vanish.

For definiteness, we now consider  the LO
expression of the  $Q_{3-6} \times Q_{3-6}$ contribution, where for
simplicity we set 
the matching coefficients $C_4, C_5$ and $C_6$ to zero and display only the terms
proportional to $C_3^2$.  Then the LO $\Delta B = 1$ amplitude including terms
of ${\cal O}(\epsilon)$ is given by
\begin{align}
  \mathcal{A}^{\Delta B = 1} &= C_3^2 \biggl[\left(14 \langle Q\rangle ^{ (0)}-25 \left\langle \tilde{Q}_S\right\rangle ^{(0)}+27 \left\langle R_0\right\rangle ^{(0)}-\frac{1}{8} \left\langle {E}_3^{ (1)}\right\rangle ^{(0)}-\frac{1}{4} \left\langle {E}_5^{ (1)}\right\rangle ^{(0)}\right) \nonumber \\
                             & + \epsilon  \left(\frac{131}{6} \langle Q\rangle ^{ (0)}-\frac{125}{3} \left\langle \tilde{Q}_S\right\rangle ^{(0)}+43 \left\langle R_0\right\rangle ^{(0)}-\frac{1}{3} \left\langle {E}_3^{ (1)}\right\rangle ^{(0)}-\frac{5}{12} \left\langle {E}_5^{ (1)}\right\rangle ^{(0)}\right) \nonumber \\
                             &+\epsilon  \left(28 \langle Q\rangle ^{ (0)}-50 \left\langle \tilde{Q}_S\right\rangle ^{(0)}+54 \left\langle R_0\right\rangle ^{(0)}-\frac{1}{4} \left\langle {E}_3^{ (1)}\right\rangle ^{(0)}-\frac{1}{2} \left\langle {E}_5^{ (1)}\right\rangle ^{(0)}\right) \log \left(\frac{{\mu_1} }{m_b}\right) \nonumber \\
                             &+\epsilon  \left(18 \langle Q\rangle ^{ (0)}-36 \left\langle
                               \tilde{Q}_S\right\rangle ^{(0)}+36 \left\langle
                               R_0\right\rangle ^{(0)}\right) z
                               + {\, {\cal O}(z^2) + {\cal O}(\epsilon^2) }
                               \biggr]\,, \label{eq:amp} 
\end{align}
where we set the number of colors to $N_c = 3$. At the same order the $\Delta B =2$ amplitude reads
\begin{align}
  \mathcal{A}^{\Delta B = 2} & = H_Q \langle Q\rangle ^{ (0)}+H_{\tilde{Q}_S} \left\langle \tilde{Q}_S\right\rangle ^{(0)}+H_{R_0} \left\langle R_0\right\rangle ^{(0)}+H_{{E}_1^{ (1)}} \left\langle {E}_1^{ (1)}\right\rangle ^{(0)}+H_{{E}_2^{ (1)}} \left\langle {E}_2^{ (1)}\right\rangle ^{(0)} \nonumber \\
                             & +H_{{E}_3^{ (1)}} \left\langle {E}_3^{ (1)}\right\rangle ^{(0)}+H_{{E}_4^{ (1)}} \left\langle {E}_4^{ (1)}\right\rangle ^{(0)}+H_{{E}_5^{ (1)}} \left\langle {E}_5^{ (1)}\right\rangle ^{(0)} + 
                               \sum_{i=2}^3 \sum_{j=1}^4 H_{E^{(i)}_j}
                    \left\langle {E}_j^{ (i)} \right \rangle{}^{(0)} \,,
\end{align}
and from the matching procedure we obtain
\begin{align}
	H_Q^{ (0,0)} &= 14 C_3^2, \nonumber\\
	H_Q^{ (0,1)} &= \frac{1}{6} C_3^2 \left(168 \log \left(\frac{\mu_1 }{m_b}\right)+108 z+131\right)\,, \nonumber\\
	H_{\tilde{Q}_S}^{ (0,0)} &= -25 C_3^2\,, \nonumber\\
	H_{\tilde{Q}_S}^{ (0,1)} &= -\frac{1}{3} C_3^2 \left(150 \log \left(\frac{\mu_1 }{m_b}\right)+108 z+125\right)\,, \nonumber\\
	H_{R_0}^{ (0,0)} &= 27 C_3^2\,, \nonumber\\
	H_{R_0}^{ (0,1)} &= C_3^2 \left(54 \log \left(\frac{\mu_1 }{m_b}\right)+36 z+43\right), \nonumber\\
	H_{{E}_3}^{ (0,0)} &= -\frac{C_3^2}{8}\,, \nonumber\\
	H_{{E}_3}^{ (0,1)} &= -\frac{1}{12} C_3^2 \left(3 \log \left(\frac{\mu_1 }{m_b}\right)+4\right)\,, \nonumber\\
	H_{{E}_5}^{ (0,0)} &= -\frac{C_3^2}{4}\,, \nonumber\\
	H_{{E}_5}^{ (0,1)} &= -\frac{1}{12} C_3^2 \left(6 \log
                             \left(\frac{\mu_1 }{m_b}\right)+5\right)\,, \label{eq:coeffs}
\end{align}
with all other $H_E^{(0,0)}$ and $H_E^{(0,1)}$ being zero. In the next step we
consider both amplitudes at NLO up to {$\mathcal{O}(\epsilon^0)$}. Upon
inserting the above values for $H_Q^{(0,0)}$, $H_Q^{(0,1)}$, $H_E^{(0,0)}$ and
$H_E^{(0,1)}$ we observe an explicit cancellation of all $1/\epsilon$ poles
multiplying $C_3^2$ which allows us to take the limit $D \to 4$. We
also find the coefficients independent of  the gauge parameter.

The presence of $R_0$ in \eqsto{eq:amp}{eq:coeffs} requires some explanation:
For $D=4$ one has $\langle R_0 \rangle^{(0)}={\cal O} (\lqcd/m_b)$ (and at NLO
and beyond $\langle R_0 \rangle ={\cal O} (\lqcd/m_b)$ is ensured by a finite
renormalization). To derive this result one employs four-dimensional Dirac
algebra (such as using the Fierz identity from \cite{Beneke:1996gn})  and for $D\neq 4$ the
definition of $R_0$ in \eq{eq:defr0} thus includes an evanescent piece.  One
may write
\begin{eqnarray}
  R_0&=& R_0^{\rm phys} + E_{R_0}
\end{eqnarray}
with $\langle R_0^{\rm phys} \rangle = {\cal O} ( \lqcd/m_b )$, while the
evanescent piece $\langle E_{R_0} \rangle$ scales as $m_b^0$.  Clearly,
if one uses a gluon mass as infra-red regulator, this subtlety does not occur,
because the matching is done in $D=4$ dimensions. In our case of
dimensional infra-red regularization, however, $E_{R_0}$ must be included in the LO
matching just as any other evanescent operator.  If we were interested
in the $C_3^2$ contributions to the $1/m_b$-suppressed part
(which is beyond the scope of this paper), we would have to provide
different coefficients for the physical operator $ R_0^{\rm phys}$ and
the unphysical $ E_{R_0}$.  For our choice of external states, namely
zero momenta $p_s$ for the light strange quarks, we cannot determine the
coefficient of $ R_0^{\rm phys} $, because
$\langle R_0^{\rm phys} \rangle^{(0)}=0$ for $p_s=0$. Therefore, the
coefficients $ H_{R_0}^{ (0,0)}$ and $H_{R_0}^{ (0,1)}$ in
\eq{eq:coeffs} are to be understood as the coefficients of $ E_{R_0}$.

The ${\cal O}(\epsilon)$ terms of the coefficients of evanescent
operators, {\it i.e.}\ $H_{{E}_3}^{(0,1)}$, $H_{{E}_5}^{(0,1)}$, and
$H_{R_0}^{ (0,1)}$ are not needed for the NLO calculation presented in
this paper. However, they will be relevant at NNLO and beyond.


\section{\label{sec::ana}Analytic results}

In this Section we present analytic results for the new
contributions to $\PPP{ab}$ and $\PPS{ab}$ introduced in Eq.~(\ref{eq::G}).
For this purpose it is convenient to decompose these quantities
according to the $|\Delta B|=1$ matching coefficients as follows
\begin{eqnarray}
  \PPP{ab} (z)  &=& {\sum_{\substack{i,j=3,\ldots, 6,8\\i\ge j}}} C_i C_j\,p_{ij}^{ab} (z)
                    \,, \nonumber\\ 
  \PPS{ab} (z)  &=& {\sum_{\substack{i,j=3,\ldots, 6,8\\i\ge j}}} C_i C_j \,p_{ij}^{S,ab} (z)
                    \,,
                    \label{eq::Hp_HSp}
\end{eqnarray}
and to write the perturbative expansion of the coefficients 
\begin{eqnarray}
  p_{ij}^{ab}(z) &=& p_{ij}^{ab,(0)}(z) + \frac{\alpha_s(\mu_1)}{4\pi}
                     p_{ij}^{ab,(1)} (z) + {\cal O}(\alpha_s^2)\,, 
  \label{eq:pp}
\end{eqnarray}
(and analogously for $p_{ij}^{S,ab}$) where $p_{ij}^{ab,(0)}$ refers
to one-loop and $p_{ij}^{ab,(1)}$ to two-loop contributions.  We
define the strong coupling constant with five active quark flavors at
the renormalization scale $\mu_1$, i.e.\ we have
$\alpha_s{(\mu_1)}\equiv\alpha_s^{(5)}(\mu_1)$. Both the charm and bottom quark
masses are defined in the on-shell scheme. Furthermore, we fix the number
of colors to $N_c=3$. Computer-readable expressions for all results
for generic $N_c$ can be downloaded from~\cite{progdata}.


\subsection{Four-quark penguin operators}

We start with the $Q_{3-6} \times Q_{3-6}$ contribution.
Both at one- and two-loop order, which contribute to
LO and NLO, respectively, the ``$cc$'', ``$uc$''
and ``$uu$'' contributions agree, because penguin operators come
with the CKM factor $-\lambda_t^s=\lambda_c^s+\lambda_u^s$:
\begin{align}
  p_{ij}^{cc,(0)}(z) &= p_{ij}^{uc,(0)}(z) = p_{ij}^{uu,(0)}(z)\,,
                       \nonumber\\
  p_{ij}^{S,cc,(0)}(z) &= p_{ij}^{S,uc,(0)}(z) = p_{ij}^{S,uu,(0)}(z)\,,
                         \nonumber\\
  p_{ij}^{cc,(1)}(z) &= p_{ij}^{uc,(1)}(z) = p_{ij}^{uu,(1)}(z)\,,
                         \nonumber\\
  p_{ij}^{S,cc,(1)}(z) &= p_{ij}^{S,uc,(1)}(z) = p_{ij}^{S,uu,(1)}(z)\,.
\end{align}
At one-loop order
exact results are available~\cite{Beneke:1996gn}, which we repeat for
convenience
\begin{eqnarray}
	p_{33}^{cc,(0)}(z) &=& \sqrt{1-4 z} \left(3 N_V+6 N_V z\right)+\left(2+3 N_L\right), \nonumber\\
	p_{34}^{cc,(0)}(z) &=& 7/3, \nonumber\\
	p_{35}^{cc,(0)}(z) &=& \sqrt{1-4 z} \left(60 N_V+120 N_V z\right)+\left(64+60 N_L\right), \nonumber\\
	p_{36}^{cc,(0)}(z) &=&  \frac{112}{3}, \nonumber\\	
	p_{44}^{cc,(0)}(z) &=&  \sqrt{1-4 z} \left(\frac{5 N_V}{12}+\frac{5 N_V z}{6}\right)+\left(\frac{13}{72}+\frac{5 N_L}{12}\right), \nonumber\\	
	p_{45}^{cc,(0)}(z) &=& \frac{112}{3} , \nonumber\\
	p_{46}^{cc,(0)}(z) &=& \sqrt{1-4 z} \left(\frac{25 N_V}{3}+\frac{50 N_V z}{3}\right)+\left(\frac{52}{9}+\frac{25 N_L}{3}\right), \nonumber\\
	p_{55}^{cc,(0)}(z) &=& \sqrt{1-4 z} \left(408 N_V-480 N_V z\right)+\left(512+408 N_L\right), \nonumber\\
	p_{56}^{cc,(0)}(z) &=& \frac{1792}{3}, \nonumber\\
	p_{66}^{cc,(0)}(z) &=& \sqrt{1-4 z} \left(\frac{170 N_V}{3}+\frac{124
                               N_V z}{3}\right)+\left(\frac{416}{9}+\frac{170
                               N_L}{3}\right), \nonumber\\
	p_{33}^{S,cc,(0)}(z) &=& \sqrt{1-4 z} \left(-6 N_V-12 N_V z\right)+\left(-1-6 N_L\right), \nonumber\\
	p_{34}^{S,cc,(0)}(z) &=& -\frac{8}{3} , \nonumber\\
	p_{35}^{S,cc,(0)}(z) &=& \sqrt{1-4 z} \left(-120 N_V-240 N_V z\right)+\left(-32-120 N_L\right), \nonumber\\
	p_{36}^{S,cc,(0)}(z) &=& -\frac{128}{3}, \nonumber\\
	p_{44}^{S,cc,(0)}(z) &=& \sqrt{1-4 z} \left(\frac{2 N_V}{3}+\frac{4 N_V z}{3}\right)+\left(-\frac{7}{9}+\frac{2 N_L}{3}\right), \nonumber\\
	p_{45}^{S,cc,(0)}(z) &=& -\frac{128}{3}, \nonumber\\
	p_{46}^{S,cc,(0)}(z) &=& \sqrt{1-4 z} \left(\frac{40 N_V}{3}+\frac{80 N_V z}{3}\right)+\left(-\frac{224}{9}+\frac{40 N_L}{3}\right), \nonumber\\
	p_{55}^{S,cc,(0)}(z) &=& \sqrt{1-4 z} \left(-816 N_V-1632 N_V z\right)+\left(-256-816 N_L\right), \nonumber\\
	p_{56}^{S,cc,(0)}(z) &=& -\frac{2048}{3}, \nonumber\\
	p_{66}^{S,cc,(0)}(z) &=& \sqrt{1-4 z} \left(\frac{272 N_V}{3}+\frac{544 N_V z}{3}\right)+\left(-\frac{1792}{9}+\frac{272 N_L}{3}\right)\,.
\end{eqnarray}
The symbols $N_L$ and $N_V$ label closed fermion loops with mass $0$ and
$m_c$, respectively. In the numerical evaluation we set $N_L=3$ and $N_V=1$.

The two-loop results are new. Their expansions up to linear order in $z$
are given by
\begin{align}
	p_{33}^{cc,(1)}(z) &= -\frac{154}{9} \logOne+\frac{16}{3} \logTwo+14 N_L \logTwo+14 N_V \logTwo +90 N_V z \nonumber \\
	&-\frac{1166}{27}+\frac{71 N_L}{3}+\frac{71 N_V}{3}+\frac{5 \pi }{3 \sqrt{3}}-\frac{5 \pi ^2}{3}, \\
	p_{34}^{cc,(1)}(z) &= -\frac{151}{54} \logOne-\frac{14}{9} N_H \logOne-\frac{8}{3} N_L \logOne-\frac{8}{3} N_V \logOne+\frac{74}{9} \logTwo-\frac{10 N_V z}{3} \nonumber \\
	& + \frac{317}{324}-\frac{5 \pi }{9 \sqrt{3}}-\frac{10 \pi ^2}{9}+N_L \left(-\frac{379}{18}+\frac{5 \pi }{3 \sqrt{3}}\right)+N_V \left(-\frac{379}{18}+\frac{5 \pi }{3 \sqrt{3}}\right) \nonumber \\
	&+N_H \left(-\frac{85}{27}+\frac{5 \pi }{3 \sqrt{3}}\right), \\
	p_{35}^{cc,(1)}(z) &=  -\frac{4928}{9} \logOne+\frac{512}{3} \logTwo+280 N_L \logTwo+280 N_V \logTwo+1800 N_V z \nonumber\\
	&-\frac{34240}{27}+\frac{1420 N_L}{3}+\frac{1420 N_V}{3}+\frac{160 \pi }{3 \sqrt{3}}-\frac{160 \pi ^2}{3}, \\
	p_{36}^{cc,(1)}(z) &= -\frac{1208}{27} \logOne-\frac{140}{9} N_H \logOne-\frac{314}{3} N_L \logOne-\frac{314}{3} N_V \logOne+144 N_V z \logOne \nonumber \\  
	&+\frac{1184}{9} \logTwo 
	+\frac{440 N_V z}{3} -\frac{13876}{81}-\frac{80 \pi }{9 \sqrt{3}}-\frac{160 \pi ^2}{9}+N_L \left(-\frac{3215}{9}+\frac{50 \pi }{3 \sqrt{3}}\right) \nonumber \\
	&+ N_V \left(-\frac{3215}{9}+\frac{50 \pi }{3 \sqrt{3}}\right) +N_H \left(-\frac{598}{27}+\frac{50 \pi }{3 \sqrt{3}}\right), \\
	p_{44}^{cc,(1)}(z) &= -\frac{187}{81} \logOne-\frac{13}{54} N_H \logOne+\frac{133}{36} N_L \logOne-\frac{5}{9} N_H N_L \logOne-\frac{5}{9} N_L^2 \logOne+\frac{133}{36} N_V \logOne \nonumber \\
	&-\frac{5}{9} N_H N_V \logOne -\frac{10}{9} N_L N_V \logOne-\frac{5}{9} N_V^2 \logOne+\frac{151}{108} \logTwo-\frac{1}{18} N_L \logTwo-\frac{1}{18} N_V \logTwo \nonumber \\ &+\left[-\frac{10}{3} N_L N_V-\frac{10 N_V^2}{3}+N_V \left(\frac{803}{36}-\frac{5 \pi ^2}{3}\right)\right] z-\frac{1466}{243}-\frac{25 N_L^2}{27} \nonumber \\  &-\frac{50 N_L N_V}{27} -\frac{25 N_V^2}{27} +\frac{5 \pi }{108 \sqrt{3}}+\frac{25 \pi ^2}{108}+N_H \left(\frac{85}{162}-\frac{5 \pi }{18 \sqrt{3}}\right) \nonumber \\  & +N_H N_L \left(-\frac{85}{27}+\frac{5 \pi }{3 \sqrt{3}}\right) +N_H N_V \left(-\frac{85}{27}+\frac{5 \pi }{3 \sqrt{3}}\right)  \nonumber \\  
	&+N_L \left(\frac{233}{27}-\frac{5 \pi }{18 \sqrt{3}}-\frac{5 \pi ^2}{6}\right) 
	+N_V \left(\frac{233}{27}-\frac{5 \pi }{18 \sqrt{3}}-\frac{5 \pi ^2}{6}\right), \\		
	p_{45}^{cc,(1)}(z) &= -\frac{1208}{27} \logOne-\frac{224}{9} N_H \logOne-\frac{362}{3} N_L \logOne-\frac{362}{3} N_V \logOne+576 N_V z \logOne \nonumber \\
	&+\frac{1184}{9} \logTwo +\frac{3836 N_V z}{3} +\frac{11234}{81}-\frac{80 \pi }{9 \sqrt{3}}-\frac{160 \pi ^2}{9} \nonumber \\
	&+N_L \left(-\frac{3754}{9}+\frac{80 \pi }{3 \sqrt{3}}\right) +N_V \left(-\frac{3754}{9}+\frac{80 \pi }{3 \sqrt{3}}\right) +N_H \left(-\frac{1360}{27}+\frac{80 \pi }{3 \sqrt{3}}\right), \\
	p_{46}^{cc,(1)}(z) &= -\frac{5984}{81} \logOne-\frac{169}{27} N_H \logOne+\frac{437}{9} N_L  \logOne-\frac{100}{9} N_H N_L  \logOne-\frac{100}{9} N_L^2  \logOne \nonumber\\
	&+\frac{437}{9} N_V \logOne -\frac{100}{9} N_H N_V \logOne-\frac{200}{9} N_L N_V \logOne-\frac{100}{9} N_V^2 \logOne+60 N_V z \logOne \nonumber \\ 
	& +\frac{1208}{27} \logTwo-\frac{10}{9} N_L \logTwo 
	-\frac{10}{9} N_V \logTwo \nonumber \\
	& +\left[-\frac{200}{3} N_L N_V-\frac{200 N_V^2}{3}+N_V \left(\frac{4855}{9}-\frac{100 \pi ^2}{3}\right)\right] z 
	\nonumber \\ 
	&-\frac{58213}{243}-\frac{410 N_L^2}{27}-\frac{820 N_L N_V}{27} -\frac{410 N_V^2}{27}+\frac{40 \pi }{27 \sqrt{3}}+\frac{200 \pi ^2}{27} \nonumber \\ 
	& +N_H N_L \left(-\frac{1610}{27}+\frac{100 \pi }{3 \sqrt{3}}\right)+N_H N_V \left(-\frac{1610}{27}+\frac{100 \pi }{3 \sqrt{3}}\right) \nonumber \\
	& +N_H \left(\frac{1222}{81}-\frac{65 \pi }{9 \sqrt{3}}\right) +N_L \left(\frac{3374}{27}-\frac{65 \pi }{9 \sqrt{3}}-\frac{50 \pi ^2}{3}\right) \nonumber \\
	& +N_V \left(\frac{3374}{27}-\frac{65 \pi }{9 \sqrt{3}}-\frac{50 \pi ^2}{3}\right), \\
	p_{55}^{cc,(1)}(z) &= -\frac{39424}{9} \logOne+\frac{4096}{3} \logTwo+1904 N_L \logTwo+1904 N_V \logTwo-2592 N_V z \logTwo \nonumber \\
	&-z \left(33120 N_V+10368 N_V \log (z)\right)-\frac{249344}{27}+\frac{16568 N_L}{3}+\frac{16568 N_V}{3} \nonumber \\
	&+\frac{1280 \pi }{3 \sqrt{3}}-\frac{1280 \pi ^2}{3}, \\		
	p_{56}^{cc,(1)}(z) &= -\frac{19328}{27} \logOne-\frac{2240}{9} N_H \logOne-\frac{5960}{3} N_L \logOne-\frac{5960}{3} N_V \logOne \nonumber \\
	&+7200 N_V z \logOne  +\frac{18944}{9} \logTwo +\frac{74000 N_V z}{3}-\frac{62560}{81}-\frac{1280 \pi }{9 \sqrt{3}}-\frac{2560 \pi ^2}{9} \nonumber \\
	&+N_L \left(-\frac{60064}{9}+\frac{800 \pi }{3 \sqrt{3}}\right)  +N_V \left(-\frac{60064}{9}+\frac{800 \pi }{3 \sqrt{3}}\right)  \nonumber \\ 
	&+N_H \left(-\frac{9568}{27}+\frac{800 \pi }{3 \sqrt{3}}\right), \\
	p_{66}^{cc,(1)}(z) &= -\frac{47872}{81} \logOne-\frac{1040}{27} N_H \logOne+\frac{2260}{9} N_L \logOne-\frac{500}{9} N_H N_L \logOne-\frac{500}{9} N_L^2 \logOne \nonumber \\
	&+\frac{2260}{9} N_V \logOne-\frac{500}{9} N_H N_V \logOne-\frac{1000}{9} N_L N_V \logOne-\frac{500}{9} N_V^2 \logOne-48 N_V z \logOne \nonumber \\
	&+\frac{9664}{27} \logTwo-\frac{68}{9} N_L \logTwo-\frac{68}{9} N_V \logTwo-144 N_V z \logTwo \nonumber \\
	& + { \left[-\frac{1000}{3} N_L N_V-\frac{1000 N_V^2}{3}+N_V \left(\frac{24290}{9}-\frac{248 \pi ^2}{3}\right)-576 N_V \log (z)\right] z} \nonumber \\
	&-\frac{556112}{243}-\frac{1600 N_L^2}{27}-\frac{3200 N_L N_V}{27}-\frac{1600 N_V^2}{27}+\frac{320 \pi }{27 \sqrt{3}}+\frac{1600 \pi ^2}{27} \nonumber \\ 
	&+N_H N_L \left(-\frac{7600}{27}+\frac{500 \pi }{3 \sqrt{3}}\right)+N_H N_V \left(-\frac{7600}{27}+\frac{500 \pi }{3 \sqrt{3}}\right) \nonumber \\ 
	&+N_H \left(\frac{8672}{81}-\frac{400 \pi }{9 \sqrt{3}}\right) +N_L \left(\frac{1148}{3}-\frac{400 \pi }{9 \sqrt{3}}-\frac{340 \pi ^2}{3}\right) \nonumber \\ 
	&+N_V \left(\frac{1148}{3}-\frac{400 \pi }{9 \sqrt{3}}-\frac{340 \pi ^2}{3}\right),
\end{align}
\begin{align}
	p_{33}^{S,cc,(1)}(z) &= \frac{176}{9} \logOne-\frac{8}{3} \logTwo-16 N_L \logTwo-16 N_V \logTwo -432 N_V z\nonumber \\
	&+\frac{1684}{27}-\frac{64 N_L}{3}-\frac{64 N_V}{3}+\frac{8 \pi }{3 \sqrt{3}}-\frac{8 \pi ^2}{3}, \\
	p_{34}^{S,cc,(1)}(z) &= \frac{220}{27} \logOne+\frac{16}{9} N_H \logOne-\frac{64}{9} \logTwo-\frac{16 N_V z}{3}+\frac{2042}{81}-\frac{8 \pi }{9 \sqrt{3}}-\frac{16 \pi ^2}{9} \nonumber \\
	&+N_H \left(-\frac{136}{27}+\frac{8 \pi }{3 \sqrt{3}}\right) 
	+N_L \left(\frac{52}{9}+\frac{8 \pi }{3 \sqrt{3}}\right)+N_V \left(\frac{52}{9}+\frac{8 \pi }{3 \sqrt{3}}\right), \\
	p_{35}^{S,cc,(1)}(z) &= \frac{5632}{9} \logOne-\frac{256}{3} \logTwo-320 N_L \logTwo-320 N_V \logTwo-8640 N_V z \nonumber \\
	&+\frac{55808}{27}-\frac{1280 N_L}{3}-\frac{1280 N_V}{3}+\frac{256 \pi }{3 \sqrt{3}}-\frac{256 \pi ^2}{3}, \\
	p_{36}^{S,cc,(1)}(z) &= \left(\frac{3520}{27} \logOne+\frac{160}{9} N_H \logOne+48 N_L \logOne+48 N_V \logOne-\frac{1024}{9} \logTwo\right)\nonumber \\
	&-\frac{160 N_V z}{3}  +\frac{47264}{81}-\frac{128 \pi }{9 \sqrt{3}}-\frac{256 \pi ^2}{9}+N_H \left(-\frac{1648}{27}+\frac{80 \pi }{3 \sqrt{3}}\right) \nonumber\\
	&+N_L \left(\frac{1288}{9}+\frac{80 \pi }{3 \sqrt{3}}\right) +N_V \left(\frac{1288}{9}+\frac{80 \pi }{3 \sqrt{3}}\right), \\
	p_{44}^{S,cc,(1)}(z) &= \frac{8}{81} \logOne+\frac{28}{27} N_H \logOne+\frac{22}{3} N_L \logOne-\frac{8}{9} N_H N_L \logOne-\frac{8}{9} N_L^2 \logOne+\frac{22}{3} N_V \logOne \nonumber\\ & -\frac{8}{9} N_H N_V \logOne 
	-\frac{16}{9} N_L N_V \logOne-\frac{8}{9} N_V^2 \logOne-\frac{56}{27} \logTwo+\frac{16}{9} N_L \logTwo+\frac{16}{9} N_V \logTwo \nonumber \\
	&+\left[-\frac{16}{3} N_L N_V-\frac{16 N_V^2}{3}+N_V \left(\frac{422}{9}-\frac{8 \pi ^2}{3}\right)\right] z \nonumber\\
	&+\frac{394}{243}-\frac{40 N_L^2}{27}-\frac{80 N_L N_V}{27}-\frac{40 N_V^2}{27}+\frac{2 \pi }{27 \sqrt{3}}+\frac{10 \pi ^2}{27} \nonumber\\
	&+N_H \left(\frac{68}{81}-\frac{4 \pi }{9 \sqrt{3}}\right) +N_H N_L \left(-\frac{136}{27}+\frac{8 \pi }{3 \sqrt{3}}\right) +N_H N_V \left(-\frac{136}{27}+\frac{8 \pi }{3 \sqrt{3}}\right)
	\nonumber \\ 
	&+N_L \left(\frac{452}{27}-\frac{4 \pi }{9 \sqrt{3}}-\frac{4 \pi ^2}{3}\right) +N_V \left(\frac{452}{27}-\frac{4 \pi }{9 \sqrt{3}}-\frac{4 \pi ^2}{3}\right), \\		
	p_{45}^{S,cc,(1)}(z) &= \frac{3520}{27} \logOne+\frac{256}{9} N_H \logOne+48 N_L \logOne+48 N_V \logOne-\frac{1024}{9} \logTwo+\frac{608 N_V z}{3} \nonumber \\
	&+\frac{26960}{81}-\frac{128 \pi }{9 \sqrt{3}}-\frac{256 \pi ^2}{9}+N_H \left(-\frac{2176}{27}+\frac{128 \pi }{3 \sqrt{3}}\right) \nonumber \\
	& +N_L \left(\frac{1232}{9}+\frac{128 \pi }{3 \sqrt{3}}\right) +N_V \left(\frac{1232}{9}+\frac{128 \pi }{3 \sqrt{3}}\right), \\
	p_{46}^{S,cc,(1)}(z) &= \frac{256}{81} \logOne+\frac{728}{27} N_H \logOne+\frac{344}{3} N_L \logOne-\frac{160}{9} N_H N_L \logOne-\frac{160}{9} N_L^2 \logOne\nonumber \\
	& +\frac{344}{3} N_V \logOne  -\frac{160}{9} N_H N_V \logOne-\frac{320}{9} N_L N_V \logOne-\frac{160}{9} N_V^2 \logOne-\frac{1792}{27} \logTwo \nonumber\\ 
	& +\frac{320}{9} N_L \logTwo +\frac{320}{9} N_V \logTwo +\biggl[-\frac{320}{3} N_L N_V-\frac{320 N_V^2}{3} \nonumber\\ 
	&+N_V \left(\frac{7408}{9}-\frac{160 \pi ^2}{3}\right)\biggr] z 
	+\frac{42680}{243}-\frac{656 N_L^2}{27}-\frac{1312 N_L N_V}{27} \nonumber\\
	& -\frac{656 N_V^2}{27}+\frac{64 \pi }{27 \sqrt{3}}+\frac{320 \pi ^2}{27} +N_H \left(\frac{1264}{81}-\frac{104 \pi }{9 \sqrt{3}}\right) \nonumber\\
	& +N_H N_L \left(-\frac{2576}{27}+\frac{160 \pi }{3 \sqrt{3}}\right) +N_H N_V \left(-\frac{2576}{27}+\frac{160 \pi }{3 \sqrt{3}}\right) \nonumber\\
	& +N_L \left(\frac{7328}{27}-\frac{104 \pi }{9 \sqrt{3}}-\frac{80 \pi ^2}{3}\right) +N_V \left(\frac{7328}{27}-\frac{104 \pi }{9 \sqrt{3}}-\frac{80 \pi ^2}{3}\right), \\
	p_{55}^{S,cc,(1)}(z) &= \frac{45056}{9} \logOne-\frac{2048}{3} \logTwo-2176 N_L \logTwo-2176 N_V \logTwo-58752 N_V z \nonumber\\
	&+\frac{461824}{27}-\frac{22528 N_L}{3}-\frac{22528 N_V}{3}+\frac{2048 \pi }{3 \sqrt{3}}-\frac{2048 \pi ^2}{3}, \\		
	p_{56}^{S,cc,(1)}(z) &= \frac{56320}{27} \logOne+\frac{2560}{9} N_H \logOne+960 N_L \logOne+960 N_V \logOne-\frac{16384}{9} \logTwo  \nonumber \\
	& +\frac{6080 N_V z}{3} +\frac{664832}{81}-\frac{2048 \pi }{9 \sqrt{3}}-\frac{4096 \pi ^2}{9}+N_H \left(-\frac{26368}{27}+\frac{1280 \pi }{3 \sqrt{3}}\right) \nonumber \\
	& +N_L \left(\frac{25472}{9}+\frac{1280 \pi }{3 \sqrt{3}}\right) +N_V \left(\frac{25472}{9}+\frac{1280 \pi }{3 \sqrt{3}}\right), \\
	p_{66}^{S,cc,(1)}(z) &= \frac{2048}{81} \logOne+\frac{4480}{27} N_H \logOne+\frac{1888}{3} N_L \logOne-\frac{800}{9} N_H N_L \logOne-\frac{800}{9} N_L^2 \logOne \nonumber\\ 
	&+\frac{1888}{3} N_V \logOne -\frac{800}{9} N_H N_V \logOne-\frac{1600}{9} N_L N_V \logOne-\frac{800}{9} N_V^2 \logOne-\frac{14336}{27} \logTwo \nonumber\\ 
	&+\frac{2176}{9} N_L \logTwo +\frac{2176}{9} N_V \logTwo +\biggl [ -\frac{1600}{3} N_L N_V-\frac{1600 N_V^2}{3} \nonumber\\ 
	&+N_V \left(\frac{42896}{9}-\frac{1088 \pi ^2}{3}\right)\biggr ] z+\frac{582016}{243}-\frac{2560 N_L^2}{27} -\frac{5120 N_L N_V}{27}\nonumber\\
	& -\frac{2560 N_V^2}{27} +\frac{512 \pi }{27 \sqrt{3}}+\frac{2560 \pi ^2}{27}+N_H \left(\frac{2816}{81}-\frac{640 \pi }{9 \sqrt{3}}\right) \nonumber\\
	& +N_H N_L \left(-\frac{12160}{27}+\frac{800 \pi }{3 \sqrt{3}}\right) +N_H N_V \left(-\frac{12160}{27}+\frac{800 \pi }{3 \sqrt{3}}\right)  \nonumber\\ 
	&+N_L \left(960-\frac{640 \pi }{9 \sqrt{3}}-\frac{544 \pi ^2}{3}\right) +N_V \left(960-\frac{640 \pi }{9 \sqrt{3}}-\frac{544 \pi ^2}{3}\right).
\end{align}
Here $N_H=1$ labels closed fermion loops with mass $m_b$ and
\begin{eqnarray}
  \logOne \, =\,  \log \frac{\mu_1^2}{m_b^2}, &&\qquad\qquad      
  \logTwo \, =\, \log \frac{\mu_2^2}{m_b^2}. \label{eq:defl12}
\end{eqnarray}  

As a novel feature compared to the NLO calculation with two current-current
operators \cite{Beneke:1998sy}, the penguin operator contributions involve
Feynman diagrams with an FCNC $b\to s$ self-energy in an external leg, 
cf. Fig.~\ref{fig::digs}. Owing
to $p_b^2=m_b^2\neq p_s^2=0$ these diagrams contribute to the result in the
same way as all other diagrams \cite{Logan:2000iv}. Indeed, we find that their
omission would lead to a divergent result.


\subsection{Chromomagnetic and four-quark operators}

In this subsection we present results for all contributions involving 
one chromomagnetic and one of the four-quark operators $Q_1, \ldots, Q_6$.
Here the one- and two-loop corrections correspond to NLO and NNLO
contributions.

We start with $Q_{1,2} \times Q_{8}$ 
where the (exact) one-loop result is given by~\cite{Beneke:1998sy}
\begin{align}
  p_{18}^{cc,(0)}(z) &= \sqrt{1-4 z} \left(\frac{5}{18}+\frac{5 z}{9}\right)\,,
                       \nonumber\\
  p_{28}^{cc,(0)}(z) &= \sqrt{1-4 z} \left(-\frac{5}{3}-\frac{10 z}{3}\right)\,,
                       \nonumber\\
  p_{18}^{{S},cc,(0)}(z) &= \sqrt{1-4 z} \left(\frac{4}{9}+\frac{8 z}{9}\right)\,,
                       \nonumber\\
  p_{28}^{{S},cc,(0)}(z) &= \left(-\frac{8}{3}-\frac{16 z}{3}\right) \sqrt{1-4 z}\,.
\end{align}
The results for $p_{i8}^{uu}$ and $p_{i8}^{S,uu}$ are obtained from
$p_{i8}^{cc}$ and $p_{i8}^{S,cc}$ for $z=0$. For $p_{ij}^{uc}$ and
$p_{ij}^{S,uc}$ we have
\begin{align}
  p^{uc,(0)}_{i8}(z) &= \frac{p^{cc,(0)}_{i8}(z) +
                       p^{uu,(0)}_{i8}}{2}\,,
                       \nonumber\\
  p^{S,uc,(0)}_{i8}(z) &= \frac{p^{S,cc,(0)}_{i8}(z) + p^{S,uu,(0)}_{i8}}{2}\,.
\end{align}

At two-loop order the results are new. The ``$cc$'' contribution is given by
\begin{align}
	p_{18}^{cc,(1)}(z) &= \left(\frac{343}{81}-\frac{5 N_H}{27}-\frac{10 N_L}{27}-\frac{10 N_V}{27}\right) \logOne-\frac{1}{27} \logTwo \nonumber \\
	&+\left(\frac{2915}{54}-\frac{10 N_L}{9}-\frac{20 N_V}{9}-\frac{10 \pi ^2}{9}\right) z  + \frac{1235}{486}-\frac{35 N_L}{81}-\frac{35 N_V}{81} \nonumber \\
	&-\frac{5 \pi }{54 \sqrt{3}} -\frac{5 \pi ^2}{9}+N_H \left(-\frac{85}{81}+\frac{5 \pi }{9 \sqrt{3}}\right),  \nonumber\\
	p_{28}^{cc,(1)}(z) &= \left(-\frac{281}{27}+\frac{10 N_H}{9}+\frac{20 N_L}{9}+\frac{20 N_V}{9}\right) \logOne+\frac{2}{9} \logTwo  \nonumber \\
	& +\left(-\frac{1133}{9}+\frac{20 N_L}{3}+\frac{40 N_V}{3}+\frac{20 \pi ^2}{3}\right) z -\frac{4475}{81}+\frac{70 N_L}{27}+\frac{70 N_V}{27} \nonumber \\
	& +\frac{5 \pi }{9 \sqrt{3}} +\frac{10 \pi ^2}{3}+N_H \left(\frac{170}{27}-\frac{10 \pi }{3 \sqrt{3}}\right)\,,
   \nonumber\\
	p_{18}^{S,cc,(1)}(z) &= \left(\frac{664}{81}-\frac{8 N_H}{27}-\frac{16 N_L}{27}-\frac{16 N_V}{27}\right) \logOne+\frac{32}{27} \logTwo \nonumber\\
	& +\left(\frac{1432}{27}-\frac{16 N_L}{9}-\frac{32 N_V}{9}-\frac{16 \pi ^2}{9}\right) z  +\frac{4660}{243}-\frac{56 N_L}{81}-\frac{56 N_V}{81} \nonumber \\
	& -\frac{4 \pi }{27 \sqrt{3}} -\frac{8 \pi ^2}{9}+N_H \left(-\frac{136}{81}+\frac{8 \pi }{9 \sqrt{3}}\right),  \nonumber\\
	p_{28}^{S,cc,(1)}(z) &= \left(-\frac{680}{27}+\frac{16 N_H}{9}+\frac{32 N_L}{9}+\frac{32 N_V}{9}\right) \logOne-\frac{64}{9} \logTwo \nonumber\\
	&+\left(-\frac{1568}{9}+\frac{32 N_L}{3}+\frac{64 N_V}{3}+\frac{32 \pi ^2}{3}\right) z  -\frac{6728}{81}+\frac{112 N_L}{27}+\frac{112 N_V}{27} \nonumber \\
	&+\frac{8 \pi }{9 \sqrt{3}} +\frac{16 \pi ^2}{3}+N_H \left(\frac{272}{27}-\frac{16 \pi }{3 \sqrt{3}}\right)
\,.
   \label{eq::12-8_cc}
\end{align}
Note that the $(uu)$ contribution is not simply obtained by taking the limit
$z\to 0$ in the expressions of Eq.~(\ref{eq::12-8_cc}) since there
are charm quark loops not connected to the external operators.
We thus have
\begin{eqnarray}
  p_{18}^{uu,(1)}(z) &=& p_{18}^{cc,(1)}(z)\Big|_{z\to 0} -\frac{10 N_V}{9} z\,,
                         \nonumber\\
  p_{28}^{uu,(1)}(z) &=& p_{28}^{cc,(1)}(z)\Big|_{z\to 0} +\frac{20 N_V }{3}z\,,
                         \nonumber\\
  p_{18}^{S,uu,(1)}(z) &=& p_{18}^{S,cc,(1)}(z)\Big|_{z\to 0} -\frac{16 N_V}{9} z\,,
                         \nonumber\\
  p_{28}^{S,uu,(1)}(z) &=& p_{28}^{S,cc,(1)}(z)\Big|_{z\to 0} +\frac{32 N_V}{3}z\,.
\end{eqnarray}
For the $uc$ contributions we find
\begin{align}
  p^{uc,(1)}_{i8}(z) &= \frac{p^{cc,(1)}_{i8}(z) +
                       p^{uu,(1)}_{i8}(z)}{2}\,,
                       \nonumber\\
  p^{S,uc,(1)}_{i8}(z) &= \frac{p^{S,cc,(1)}_{i8}(z) +
                         p^{S,uu,(1)}_{i8}(z)}{2}
                         \,.
\end{align}

For the contribution $Q_{3-6} \times Q_{8}$ we observe that
both at one- and two-loop order we obtain the same results
for the ``$cc$'', ``$uu$'' and ``$uc$'' contributions and thus
we have
\begin{eqnarray}
  p_{ij}^{cc,(0)}(z) &=& p_{ij}^{uc,(0)}(z) = p_{ij}^{uu,(0)}(z) \,,
                         \nonumber\\
  p_{ij}^{S,cc,(0)}(z) &=& p_{ij}^{S,uc,(0)}(z) = p_{ij}^{S,uu,(0)}(z)\,,
                         \nonumber\\
  p_{ij}^{cc,(1)}(z) &=& p_{ij}^{uc,(1)}(z) = p_{ij}^{uu,(1)}(z)\,,
                         \nonumber\\
  p_{ij}^{S,cc,(1)}(z) &=& p_{ij}^{S,uc,(1)}(z) = p_{ij}^{S,uu,(1)}(z)\,.
\end{eqnarray}

The one-loop results are exact in $z$ and read
\begin{align}
	p_{38}^{cc,(0)}(z) &= -\frac{32}{3} , \nonumber\\
	p_{48}^{cc,(0)}(z) &= \sqrt{1-4 z} \left(-\frac{5 N_V}{3}-\frac{10 N_V z}{3}\right)+\left(-\frac{49}{18}-\frac{5 N_L}{3}\right), \nonumber\\
	p_{58}^{cc,(0)}(z) &= -\frac{512}{3}, \nonumber\\
	p_{68}^{cc,(0)}(z) &= \sqrt{1-4 z} \left(-\frac{50 N_V}{3}-\frac{100 N_V z}{3}\right)+\left(-\frac{392}{9}-\frac{50 N_L}{3}\right),
                             \nonumber\\
	p_{38}^{S,cc,(0)}(z) &= \frac{64}{3} , \nonumber\\
	p_{48}^{S,cc,(0)}(z) &= \sqrt{1-4 z} \left(-\frac{8 N_V}{3}-\frac{16 N_V z}{3}\right)+\left(\frac{76}{9}-\frac{8 N_L}{3}\right), \nonumber\\
	p_{58}^{S,cc,(0)}(z) &= \frac{1024}{3}, \nonumber\\
	p_{68}^{S,cc,(0)}(z) &= \sqrt{1-4 z} \left(-\frac{80 N_V}{3}-\frac{160 N_V z}{3}\right)+\left(\frac{1216}{9}-\frac{80 N_L}{3}\right)\,.
\end{align}
At two-loop order our results read
\begin{align}
	p_{38}^{cc,(1)}(z) &= -\frac{1285}{27} \logOne+\frac{64}{9} N_H \logOne+\frac{28}{3} N_L \logOne+\frac{28}{3} N_V \logOne-\frac{448}{9} \logTwo -\frac{196 N_V z}{3} \nonumber \\
	&-\frac{30707}{81}+\frac{193 \pi }{18 \sqrt{3}}+\frac{25 \pi ^2}{6}+N_H \left(\frac{170}{27}-\frac{10 \pi }{3 \sqrt{3}}\right)+N_L \left(\frac{361}{9}-\frac{10 \pi }{3 \sqrt{3}}\right)  \nonumber \\
	&+N_V \left(\frac{361}{9}-\frac{10 \pi }{3 \sqrt{3}}\right), \\
	p_{48}^{cc,(1)}(z) &= -\frac{1469}{162} \logOne+\frac{98}{27} N_H \logOne-\frac{799}{54} N_L \logOne+\frac{20}{9} N_H N_L \logOne +\frac{20}{9} N_L^2 \logOne \nonumber \\
	&-\frac{799}{54} N_V \logOne +\frac{20}{9} N_H N_V \logOne+\frac{40}{9} N_L N_V \logOne+\frac{20}{9} N_V^2 \logOne-\frac{451}{27} \logTwo+\frac{2}{9} N_L \logTwo \nonumber \\
	&+\frac{2}{9} N_V \logTwo +\left[\frac{40 N_L N_V}{3}+\frac{40 N_V^2}{3}+N_V \left(-188+\frac{20 \pi ^2}{3}\right)\right] z-\frac{41707}{486}\nonumber \\
	&+\frac{100 N_L^2}{27}+\frac{200 N_L N_V}{27}+\frac{100 N_V^2}{27}-\frac{841 \pi }{108 \sqrt{3}}+\frac{17 \pi ^2}{36}+N_H N_L \left(\frac{340}{27}-\frac{20 \pi }{3 \sqrt{3}}\right)\nonumber \\
	&+N_H N_V \left(\frac{340}{27}-\frac{20 \pi }{3 \sqrt{3}}\right) +N_H \left(-\frac{3695}{162}+\frac{395 \pi }{36 \sqrt{3}}+\frac{5 \pi ^2}{18}\right)\nonumber \\
	&+N_L \left(-\frac{3605}{81}+\frac{10 \pi }{9 \sqrt{3}}+\frac{10 \pi ^2}{3}\right)+N_V \left(-\frac{3605}{81}+\frac{10 \pi }{9 \sqrt{3}}+\frac{10 \pi ^2}{3}\right), \\
	p_{58}^{cc,(1)}(z) &= -\frac{20560}{27} \logOne+\frac{1024}{9} N_H \logOne+\frac{628}{3} N_L \logOne+\frac{628}{3} N_V \logOne-\frac{7168}{9} \logTwo-\frac{760 N_V z}{3}\nonumber \\
	&-\frac{540206}{81}+\frac{1940 \pi }{9 \sqrt{3}}+\frac{578 \pi ^2}{9}+N_L \left(\frac{3476}{9}-\frac{160 \pi }{3 \sqrt{3}}\right)+N_V \left(\frac{3476}{9}-\frac{160 \pi }{3 \sqrt{3}}\right)\nonumber \\
	&+N_H \left(-\frac{5056}{27}-\frac{16 \pi }{3 \sqrt{3}}+\frac{64 \pi ^2}{3}\right), \\
	p_{68}^{cc,(1)}(z) &= -\frac{11752}{81} \logOne+\frac{1274}{27} N_H \logOne-\frac{3086}{27} N_L \logOne+\frac{200}{9} N_H N_L \logOne+\frac{200}{9} N_L^2 \logOne \nonumber \\
	&-\frac{3086}{27} N_V \logOne+\frac{200}{9} N_H N_V \logOne+\frac{400}{9} N_L N_V \logOne+\frac{200}{9} N_V^2 \logOne-\frac{7216}{27} \logTwo     \nonumber\\ 
	&+\frac{20}{9} N_L \logTwo+\frac{20}{9} N_V \logTwo\nonumber +\left[\frac{400 N_L N_V}{3}+\frac{400 N_V^2}{3}+N_V \left(-\frac{5822}{3}+\frac{200 \pi ^2}{3}\right)\right] z\nonumber \\
	&-\frac{249917}{243}+\frac{820 N_L^2}{27}+\frac{1640 N_L N_V}{27}+\frac{820 N_V^2}{27}-\frac{970 \pi }{27 \sqrt{3}}+\frac{71 \pi ^2}{27}\nonumber \\
	&+N_H N_L \left(\frac{3220}{27}-\frac{200 \pi }{3 \sqrt{3}}\right) +N_H N_V \left(\frac{3220}{27}-\frac{200 \pi }{3 \sqrt{3}}\right)\nonumber \\
	&+N_H \left(-\frac{22297}{81}+\frac{1130 \pi }{9 \sqrt{3}}+\frac{10 \pi ^2}{3}\right)+N_L \left(-\frac{32654}{81}+\frac{130 \pi }{9 \sqrt{3}}+\frac{100 \pi ^2}{3}\right) \nonumber \\
	&+N_V \left(-\frac{32654}{81}+\frac{130 \pi }{9 \sqrt{3}}+\frac{100 \pi ^2}{3}\right) , \\
	p_{38}^{S,cc,(1)}(z) &= \frac{1976}{27} \logOne-\frac{128}{9} N_H \logOne-\frac{32}{3} N_L \logOne-\frac{32}{3} N_V \logOne+\frac{512}{9} \logTwo +\frac{608 N_V z}{3} \nonumber \\
	&+\frac{27160}{81}+\frac{188 \pi }{9 \sqrt{3}}-\frac{596 \pi ^2}{27}+N_L \left(-\frac{152}{9}-\frac{16 \pi }{3 \sqrt{3}}\right)+N_V \left(-\frac{152}{9}-\frac{16 \pi }{3 \sqrt{3}}\right) \nonumber \\
	& +N_H \left(\frac{272}{27}-\frac{16 \pi }{3 \sqrt{3}}\right), \\
	p_{48}^{S,cc,(1)}(z) &= \frac{3548}{81} \logOne-\frac{304}{27} N_H \logOne-\frac{1100}{27} N_L \logOne+\frac{32}{9} N_H N_L \logOne+\frac{32}{9} N_L^2 \logOne \nonumber \\
	& -\frac{1100}{27} N_V \logOne +\frac{32}{9} N_H N_V \logOne+\frac{64}{9} N_L N_V \logOne+\frac{32}{9} N_V^2 \logOne+\frac{608}{27} \logTwo \nonumber \\
	&-\frac{64}{9} N_L \logTwo-\frac{64}{9} N_V \logTwo  +\left[\frac{64 N_L N_V}{3}+\frac{64 N_V^2}{3}+N_V \left(-128+\frac{32 \pi ^2}{3}\right)\right] z \nonumber\\
	&+\frac{38584}{243}+\frac{160 N_L^2}{27}+\frac{320 N_L N_V}{27}+\frac{160 N_V^2}{27}-\frac{634 \pi }{27 \sqrt{3}}-\frac{674 \pi ^2}{81} \nonumber \\
	&+N_H N_L \left(\frac{544}{27}-\frac{32 \pi }{3 \sqrt{3}}\right)+N_H N_V \left(\frac{544}{27}-\frac{32 \pi }{3 \sqrt{3}}\right) \nonumber \\
	& +N_H \left(-\frac{2956}{81}+\frac{158 \pi }{9 \sqrt{3}}+\frac{4 \pi ^2}{9}\right) +N_L \left(-\frac{9944}{81}+\frac{16 \pi }{9 \sqrt{3}}+\frac{16 \pi ^2}{3}\right) \nonumber\\
	& +N_V \left(-\frac{9944}{81}+\frac{16 \pi }{9 \sqrt{3}}+\frac{16 \pi ^2}{3}\right), \\
	p_{58}^{S,cc,(1)}(z) &= \left(\frac{31616}{27} \logOne-\frac{2048}{9} N_H \logOne-\frac{224}{3} N_L \logOne-\frac{224}{3} N_V \logOne+\frac{8192}{9} \logTwo\right)\nonumber\\
	&+\frac{11456 N_V z}{3}+\frac{502864}{81}+\frac{2720 \pi }{9 \sqrt{3}}-\frac{9488 \pi ^2}{27} +N_L \left(-\frac{2656}{9}-\frac{256 \pi }{3 \sqrt{3}}\right) \nonumber\\
	&+N_V \left(-\frac{2656}{9}-\frac{256 \pi }{3 \sqrt{3}}\right)+N_H \left(\frac{4352}{27}-\frac{544 \pi }{3 \sqrt{3}}+\frac{64 \pi ^2}{3}\right), \\
	p_{68}^{S,cc,(1)}(z) &= \frac{56768}{81} \logOne-\frac{3952}{27} N_H \logOne-\frac{10928}{27} N_L \logOne+\frac{320}{9} N_H N_L \logOne+\frac{320}{9} N_L^2 \logOne \nonumber \\
	&-\frac{10928}{27} N_V \logOne+\frac{320}{9} N_H N_V \logOne+\frac{640}{9} N_L N_V \logOne+\frac{320}{9} N_V^2 \logOne+\frac{9728}{27} \logTwo \nonumber \\
	&-\frac{640}{9} N_L \logTwo-\frac{640}{9} N_V \logTwo +\biggl[\frac{640 N_L N_V}{3}+\frac{640 N_V^2}{3} +N_V \left(-\frac{2720}{3}+\frac{320 \pi ^2}{3}\right)\biggr] z \nonumber \\
	&+\frac{458776}{243}+\frac{1312 N_L^2}{27}+\frac{2624 N_L N_V}{27}+\frac{1312 N_V^2}{27}-\frac{4816 \pi }{27 \sqrt{3}}-\frac{11672 \pi ^2}{81}  \nonumber \\
	& +N_H N_L \left(\frac{5152}{27}-\frac{320 \pi }{3 \sqrt{3}}\right) +N_H N_V \left(\frac{5152}{27}-\frac{320 \pi }{3 \sqrt{3}}\right)  \nonumber \\
	&+N_H \left(-\frac{27640}{81}+\frac{1808 \pi }{9 \sqrt{3}}\right)+N_L \left(-\frac{97808}{81}+\frac{208 \pi }{9 \sqrt{3}}+\frac{160 \pi ^2}{3}\right) \nonumber \\
	&+N_V \left(-\frac{97808}{81}+\frac{208 \pi }{9 \sqrt{3}}+\frac{160 \pi ^2}{3}\right) ,
\end{align}


\subsection{Two chromomagnetic operators}

Finally, we come to the  $Q_{8} \times Q_{8}$ contribution, where the one-loop
corrections are already of NNLO.
The one-loop result, for which only the $N_f$-piece has been known in the literature, is given by
\begin{align}
  p_{88}^{cc,(0)} (z) &= p_{88}^{uc,(0)} (z) = p_{88}^{uu,(0)} (z) =
                    -\frac{133}{18}+\frac{5 N_L}{3}+ \sqrt{1-4 z} \left
                    (\frac{5}{3} N_V +\frac{10}{3} N_V  z \right )\,,
                    \nonumber   \\  
  p_{88}^{S,cc,(0)} (z) &= p_{88}^{S,uc,(0)} (z)
                      = p_{88}^{S,uu,(0)} (z)
                      = -\frac{164}{9}+\frac{8 N_L}{3}+ \sqrt{1-4 z} \left (
                      \frac{8}{3} N_V +\frac{16}{3} N_V  z \right )
                      \,.
\end{align}
At two-loop order we have
\begin{align}
  p_{88}^{cc,(1)} &= p_{88}^{uc,(1)} = p_{88}^{uu,(1)}\,, 
      \nonumber              \\
  p_{88}^{S,cc,(1)}(z) &= p_{88}^{S,uc,(1)}(z) = p_{88}^{S,uu,(1)}(z)\,,
\end{align}
with
\begin{align}
	p_{88}^{cc,(1)} &= \biggl(-\frac{2527}{27}+\frac{266 N_H}{27}+\frac{836 N_L}{27}-\frac{20 N_H N_L}{9}-\frac{20 N_L^2}{9}+\frac{836 N_V}{27}-\frac{20 N_H N_V}{9} \nonumber \\
	&-\frac{40 N_L N_V}{9} -\frac{20 N_V^2}{9}\biggr) \logOne+\left(\frac{257}{27}-\frac{2 N_L}{9}-\frac{2 N_V}{9}\right) \logTwo+\biggl[-\frac{40}{3} N_L N_V \nonumber \\
	& -\frac{40 N_V^2}{3}+N_V \left(\frac{853}{9}-\frac{20 \pi ^2}{3}\right)\biggr ] z -\frac{156295}{486}-\frac{100 N_L^2}{27}-\frac{200 N_L N_V}{27}-\frac{100 N_V^2}{27} \nonumber\\
	&+\frac{277 \pi }{18 \sqrt{3}}+\frac{167 \pi ^2}{27} +N_H N_L \left(-\frac{340}{27}+\frac{20 \pi }{3 \sqrt{3}}\right) +N_H N_V \left(-\frac{340}{27}+\frac{20 \pi }{3 \sqrt{3}}\right)\nonumber\\
	&+N_L \left(\frac{8632}{81}-\frac{10 \pi }{9 \sqrt{3}}-\frac{10 \pi ^2}{3}\right) +N_V \left(\frac{8632}{81}-\frac{10 \pi }{9 \sqrt{3}}-\frac{10 \pi ^2}{3}\right) \nonumber \\
	&+N_H \left(\frac{1175}{27}-\frac{125 \pi }{6 \sqrt{3}}-\frac{5 \pi ^2}{9}\right), \\
	p_{88}^{S,cc,(1)} &= \biggl(-\frac{6232}{27}+\frac{656 N_H}{27}+\frac{1568 N_L}{27}-\frac{32 N_H N_L}{9}-\frac{32 N_L^2}{9}+\frac{1568 N_V}{27}-\frac{32 N_H N_V}{9} \nonumber \\
	&-\frac{64 N_L N_V}{9}-\frac{32 N_V^2}{9}\biggr) \logOne+\left(-\frac{1312}{27}+\frac{64 N_L}{9}+\frac{64 N_V}{9}\right) \logTwo \nonumber \\
	&+\left[-\frac{64}{3} N_L N_V-\frac{64 N_V^2}{3}+N_V \left(\frac{616}{9}-\frac{32 \pi ^2}{3}\right)\right] z -\frac{222200}{243}-\frac{160 N_L^2}{27} \nonumber \\
	&-\frac{320 N_L N_V}{27}-\frac{160 N_V^2}{27}+\frac{140 \pi }{3 \sqrt{3}}+\frac{1828 \pi ^2}{81}+N_H N_L \left(-\frac{544}{27}+\frac{32 \pi }{3 \sqrt{3}}\right) \nonumber\\
	& +N_H N_V \left(-\frac{544}{27}+\frac{32 \pi }{3 \sqrt{3}}\right)+N_L \left(\frac{15856}{81}-\frac{16 \pi }{9 \sqrt{3}}-\frac{16 \pi ^2}{3}\right) \nonumber \\
	&+N_V \left(\frac{15856}{81}-\frac{16 \pi }{9 \sqrt{3}}-\frac{16 \pi ^2}{3}\right) +N_H \left(\frac{1880}{27}-\frac{100 \pi }{3 \sqrt{3}}-\frac{8 \pi ^2}{9}\right)\,.
\end{align}



\section{\label{sec::num}Numerical results}

In this section we present the numerical effect of the new corrections
to $\Delta\Gamma_s$ and $a_{\rm fs}^s$. We start with discussing the
relative size of the contributions from the various operators and
consider afterwards the ratio $\Delta\Gamma_s/\Delta M_s$, from which
$|V_{ts}|$ and the ballpark of the hadronic uncertainties cancel.
Finally, we use the measured result for $\Delta M_s$ and present updated
results for $\Delta\Gamma_s$ in two different renormalization
schemes. We also present updated results for $a_{\rm fs}^s$.

The calculations described in the previous sections and the analytic results
presented in Section~\ref{sec::ana} use the $\overline{\rm MS}$ scheme for the
strong coupling constant and the operator mixing and the on-shell scheme for
the charm and bottom quark masses. It is well known that the latter choice
leads to large perturbative corrections. Thus, we choose as our default
renormalization scheme the one where all parameters are defined in the
$\overline{\rm MS}$ scheme. It is obtained with the help of the one-loop
relations between the on-shell and $\overline{\rm MS}$ charm and bottom quark
masses.  We define a second renormalization scheme where the overall factor
$m_b^2$ (see, e.g., \eq{eq::Gam^ab}) is defined in the on-shell
scheme, but $H^{ab}$ and $\tilde{H}_S^{ab}$ depend on the quark masses in the
$\overline{\rm MS}$ scheme. In the following we refer to this scheme as the ``pole''
scheme~\cite{Asatrian:2017qaz,Asatrian:2020zxa}.  Note that after each scheme
change, which adds $z$-exact expressions to the two-loop term, we re-expand
the latter in $z$ up to linear order to be consistent with our genuine
two-loop calculation.

\begin{table}[t]
  \begin{center}
    \begin{tabular}{rclc}
      \hline 
      $\alpha_s(M_Z)$ &=& $0.1179 \pm 0.001$ & \cite{Zyla:2020zbs}
      \\
      $m_c(3~\mbox{GeV})$ &=& $0.993\pm 0.008$~GeV & \cite{Chetyrkin:2017lif}
      \\
      $m_b(m_b)$ &=& $4.163\pm 0.016$~GeV & \cite{Chetyrkin:2017lif} 
      \\
      $m_t^{\rm pole}$ &=& $172.9\pm 0.4$~\mbox{GeV} & \cite{Zyla:2020zbs} 
      \\
      $M_{B_s}$ &=& $5366.88$~\mbox{MeV} & \cite{Zyla:2020zbs} 
      \\
      $B_{B_s}$ &=& $0.813\pm0.034$ & \cite{Dowdall:2019bea} 
      \\
      $\tilde{B}^\prime_{S,B_s}$ &=& $1.31\pm0.09$ & \cite{Dowdall:2019bea} 
      \\
      $f_{B_s}$ &=& $0.2307\pm0.0013$~GeV & \cite{Bazavov:2017lyh} 
      \\
      \hline 
    \end{tabular}
  \end{center}
  \caption{\label{tab::input}Input parameters for the numerical analysis.
    From the charm and bottom quark mass one obtains
    $\bar z = 0.04974 \pm 0.00092$. The quoted $m_t^{\rm pole}$ corresponds to
    $m_t(m_t)=(163.1\pm 0.4)\,\gev$ in the $\ov{\rm MS}$ scheme. We use
    the values for $B_{B_s}=B_{B_s}(\mu_2)$ and
    $\tilde{B}^\prime_{S,B_s}=\tilde{B}^\prime_{S,B_s}(\mu_2)$ with
    $\mu_2=m_b^{\rm pole}=4.56\,\gev$.  }
    ~\\[-4mm] \hrule
\end{table}

For convenience, we summarize in Tab.~\ref{tab::input} the input parameters
needed for our numerical analysis. In addition we have (see Ref.~\cite{Asatrian:2020zxa})
\begin{eqnarray}
  \frac{\lambda^s_u}{\lambda^s_t} &=& -(0.00865 \pm 0.00042) +(0.01832 \pm 0.00039)i\,.
                                      \label{eq::ckm_input}
\end{eqnarray}
From $m_b(m_b)$ we obtain $m_b^{\rm pole}=4.56$~GeV using the
one-loop conversion formula.
$B_{B_s}$ and $\tilde{B}^\prime_{S,B_s}$ parametrize the matrix elements
of $Q$ and $\tilde{Q}_S$ as
\begin{eqnarray}
 \bra{B_s} Q (\mu_2) \ket{\ov B_s}  &=&
   \frac{8}{3} M^2_{B_s}\, f^2_{B_s} B_{B_s} (\mu_2),
\nonumber \\
\bra{B_s} \tilde Q_S (\mu_2)\ket{\ov B_s} &=& \frac{1}{3}  M^2_{B_s}\,
  f^2_{B_s} \tilde  B_{S,B_s}^\prime (\mu_2).
      \label{eq:defb}
\end{eqnarray}  
For the matrix elements of the $1/m_b$
suppressed corrections we have
\begin{align}
  \braket{B_s|R_0|\bar{B}_s} &= - (0.43 \pm 0.17) f_{B_s}^2 M_{B_s}^2\,,
  \nonumber\\
  \braket{B_s|R_1|\bar{B}_s} &=  (0.07 \pm 0.00) f_{B_s}^2 M_{B_s}^2\,,
  \nonumber\\
  \braket{B_s|\tilde{R}_1|\bar{B}_s} &= (0.04 \pm 0.00)
  f_{B_s}^2 M_{B_s}^2\,, \nonumber\\
  \braket{B_s|R_2|\bar{B}_s} &= - (0.18 \pm 0.07) f_{B_s}^2 M_{B_s}^2\,,
  \nonumber\\
  \braket{B_s|\tilde{R}_2|\bar{B}_s} &=  (0.18 \pm 0.07) f_{B_s}^2
  M_{B_s}^2\,, \nonumber\\ 
  \braket{B_s|R_3|\bar{B}_s} &=  (0.38 \pm 0.13) f_{B_s}^2 M_{B_s}^2\,,
  \nonumber\\
  \braket{B_s|\tilde{R}_3|\bar{B}_s} &=  (0.29 \pm 0.10) f_{B_s}^2
                                       M_{B_s}^2\,.	 
  \label{eq::1/mb_ME}
\end{align}
The results for $\braket{B_s|R_2|\bar{B}_s}$,
$\braket{B_s|\tilde{R}_2|\bar{B}_s}$, $\braket{B_s|R_3|\bar{B}_s}$, and
$\braket{B_s|\tilde{R}_3|\bar{B}_s}$ can be found in
Ref.~\cite{Davies:2019gnp} and we extract the remaining three matrix elements
from~\cite{Dowdall:2019bea}. For $\braket{B_s|R_1|\bar{B}_s}$ and
$\braket{B_s|\tilde{R}_1|\bar{B}_s}$ the ratio of the bottom and strange quark
masses is needed 
${m_b(\mu)/m_s(\mu)} = 52.55 \pm 0.55$~\cite{Chakraborty:2014aca}.  

Let us next discuss our choices for the various renormalization schemes.
We fix the high scale in the $\Delta B=1$ theory to
$\mu_0=165~\mbox{GeV}\approx 2 m_W \approx m_t(m_t)$.  Since $\mu_2$
is closely connected to the lattice results for $B_{B_q}$,
$\tilde{B}^\prime_{S,B_s}$ and the $1/m_b$ matrix elements of
Eq.~(\ref{eq::1/mb_ME}), we fix it to $\mu_2=m_b^{\rm pole}$.
For $\mu_1$ we choose $m_b(m_b)$ and $m_b^{\rm pole}$ in the
$\overline{\rm MS}$ and pole renormalization scheme, respectively.
Furthermore, there are the renormalization scales $\mu_c$ and $\mu_b$ of
the charm and bottom quark masses, which in principle can be varied
independently. However, choosing $\mu_c=\mu_b$ avoids potentially large
logarithms $z\log z$ \cite{Beneke:2002rj} which is why our default
choice is $\mu_c=\mu_b=m_b(m_b)$. That is, instead of
  $z = (m_c^{\rm pole}/m_b^{\rm pole})^2$ we use
\begin{eqnarray}
 \bar z  &=& \frac{m_c^2(\mu_b)}{m_b^2(\mu_b)} \nonumber
\end{eqnarray}
as in \cite{Beneke:2002rj,Beneke:2003az,Lenz:2006hd,
  Asatrian:2017qaz,Asatrian:2020zxa,Gerlach:2021xtb}.  This means that
the coefficients $p_{ij}^{ab,(1)}(z)$ and $p_{ij}^{S,ab,(1)}(z)$ must be
replaced by $\bar p_{ij}^{ab,(1)}({\bar z})$ and
$\bar p_{ij}^{S,ab,(1)}(\bar z)$, respectively, as defined in Eq.~(32)
of Ref.~\cite{Gerlach:2021xtb}.

\begin{table}[t]
  \begin{center}
    \renewcommand{\arraystretch}{1.2}
    {\scalefont{0.8}
    \begin{tabular}{rcl|r|r|c}
      \hline 
      \multicolumn{3}{c|}{Contribution $X$} 
      & \multicolumn{1}{c|}{$r_X$ ($\overline{\rm MS}$)}
      & \multicolumn{1}{c|}{$r_X$ (pole)} 
      \\
      \hline
      $Q_{1,2}$ &$\times$& $Q_{1,2}$ & 133 (145, $-$12.0)\%        & 141 (190, $-$49.2) \% &(LO,NLO) \\
      $Q_{1,2}$ &$\times$& $Q_{3-6}$ & $-$9.55 ($-$9.02, $-$0.53)\%& $-$9.82 ($-$11.5, 1.63)\% &(LO,NLO) \\
      $Q_{3-6}$ &$\times$& $Q_{3-6}$ & 1.67 (1.32, 0.35)\%         & 1.74 (1.60, 0.14)\% &(LO,NLO) \\
      $Q_{1,2}$ &$\times$& $Q_{8}$   & 1.01 (0.78, 0.23)\%         & 1.09 (0.98, 0.11)\% &(NLO,NNLO) \\
      $Q_{3,6}$ &$\times$& $Q_{8}$   & $-$0.33 ($-$0.21, $-$0.12)\%& $-$0.36 ($-$0.26, $-$0.09)\% &(NLO,NNLO) \\
      $Q_{8}$   &$\times$& $Q_{8}$   & $-$0.33 ($-$0.20, $-$0.12) $10^{-2}$ \%
                                                                   & $-$0.36 ($-$0.25, $-$0.11) $10^{-2}$ \% &(NNLO,N$^3$LO) \\
      \hline 
    \end{tabular} }
\end{center}
    \caption{\label{tab::rel_contr}Relative contributions in percent in the
      $\overline{\rm MS}$ and pole schemes. The breakdown into one- and two-loop
      contributions is shown inside the round brackets. In the last column we
      mention the corresponding perturbative order.  }
      ~\\[-4mm] \hrule
\end{table}

In Tab.~\ref{tab::rel_contr} we show the relative size of the individual
contributions to $\Delta\Gamma_s$ both in the $\overline{\rm MS}$
and pole scheme. They are defined as
\begin{eqnarray}
  r_X &=& \frac{\Delta\Gamma_s^X}{\Delta\Gamma_s}
          \,,
          \label{eq::rX}
\end{eqnarray}
with
$X\in\{Q_{1,2}\times Q_{1,2}, Q_{1,2}\times Q_{3-6}, Q_{3-6}\times Q_{3-6},
\ldots\}$. Power-suppressed $1/m_b$ corrections are only included in the
denominator of Eq.~(\ref{eq::rX}) but not in the numerator.  In both
renormalization schemes the dominant contribution is given by the
$Q_{1,2}\times Q_{1,2}$, followed about a $7\%$ contribution from
$Q_{1,2}\times Q_{3-6}$. The remaining terms contribute at the 1\% level or
below. Note that these contributions are necessary to obtain complete NLO and
NNLO corrections. It is interesting to note that the QCD corrections to
$Q_{1,2}\times Q_{1,2}$ amount only to 9\% in the $\overline{\rm MS}$ scheme
but to more than 30\% in the pole scheme.  Also for the contribution
$Q_{1,2}\times Q_{3-6}$ the QCD corrections are about a factor of three larger
in the pole scheme whereas for $Q_{3-6}\times Q_{3-6}$ the situation is vice
versa.  For the contributions involving $Q_8$ the QCD corrections in the
$\overline{\rm MS}$ scheme amount to up to about 50\% of the leading
order term, though their absolute contribution is small.

Let us next consider $\Delta\Gamma_s/\Delta M_s$. We use
Eq.~(\ref{eq:dgdm}) with $\Gamma_{12}^s$ from Eq.~(\ref{eq:ot}) and
$M_{12}^s$ from Ref.~\cite{Buras:1990fn} where two-loop QCD corrections have
been computed.   In the two renormalization schemes our results
read
\begin{eqnarray}
  \frac{\Delta\Gamma_s}{\Delta M_s} 
  &=&
  (4.70 
      {}^{+0.32}_{-0.70}{}_{\rm scale}
      \pm 0.12_{B\tilde{B}_S}
      \pm 0.80_{1/m_b}
      \pm 0.05_{\textrm{input}})
      \times 10^{-3}
      \quad
  (\textrm{pole})\,,  \nonumber\\
  \frac{\Delta\Gamma_s}{\Delta M_s} 
  &=&  
  (5.20
      {}^{+0.01}_{-0.16}{}_{\rm scale}
      \pm 0.12_{B\tilde{B}_S}
      \pm 0.67_{1/m_b}
      \pm 0.06_{\textrm{input}}) \times 10^{-3} \quad
  (\overline{\textrm{MS}})\,, 
  \label{eq::dGdM}
\end{eqnarray}
where the subscripts indicate the source of the uncertainties: ``scale''
denotes the uncertainties from the variation of $\mu_1$,
``$B\tilde{B}_S$'' those from the leading order bag parameters and
``input'' refers to the variation of $\alpha_s(m_Z)$, $m_b(m_b)$,
$m_c(3\textrm{ GeV})$, $m_t^{\rm pole}$ and the CKM parameters in
Eq.~(\ref{eq::ckm_input}).  The uncertainties from the matrix elements
of the power-suppressed corrections in Eq.~(\ref{eq::1/mb_ME}) are
denoted by ``$1/m_b$''.  Adding the uncertainties in quadrature 
(and symmetrising the scale uncertainty) yields the numbers quoted in the
abstract.

The largest uncertainty is induced by the power-suppressed $1/m_b$
corrections. It is obtained by combining the uncertainties from the seven
matrix elements of Eq.~(\ref{eq::1/mb_ME}) in quadrature taking into account
the 100\% correlation of $\braket{B_s|R_2|\bar{B}_s}$ and 
$\braket{B_s|\tilde{R}_2|\bar{B}_s}$.  Next, there is the renormalization
scale uncertainty, which we use to estimate the contribution from unknown
higher order corrections. We obtain the numbers in Eq.~(\ref{eq::dGdM}) by
varying $\mu_1$ between $2.5\textrm{ GeV}$ and $10.0\textrm{ GeV}$ while
keeping $\mu_2$, $\mu_c$ and $\mu_b$ at their default values. A simultaneous
variation of $\mu_1 = \mu_b = \mu_c$ leads to significant larger scale
uncertainties, which is expected, because the anomalous dimension of the quark
mass is large and appears in the coefficient of $\log(\mu_b/m_b)$.

\begin{figure}[t]
  \begin{center}
    \begin{tabular}{cc}
      \includegraphics[width=0.48\textwidth]{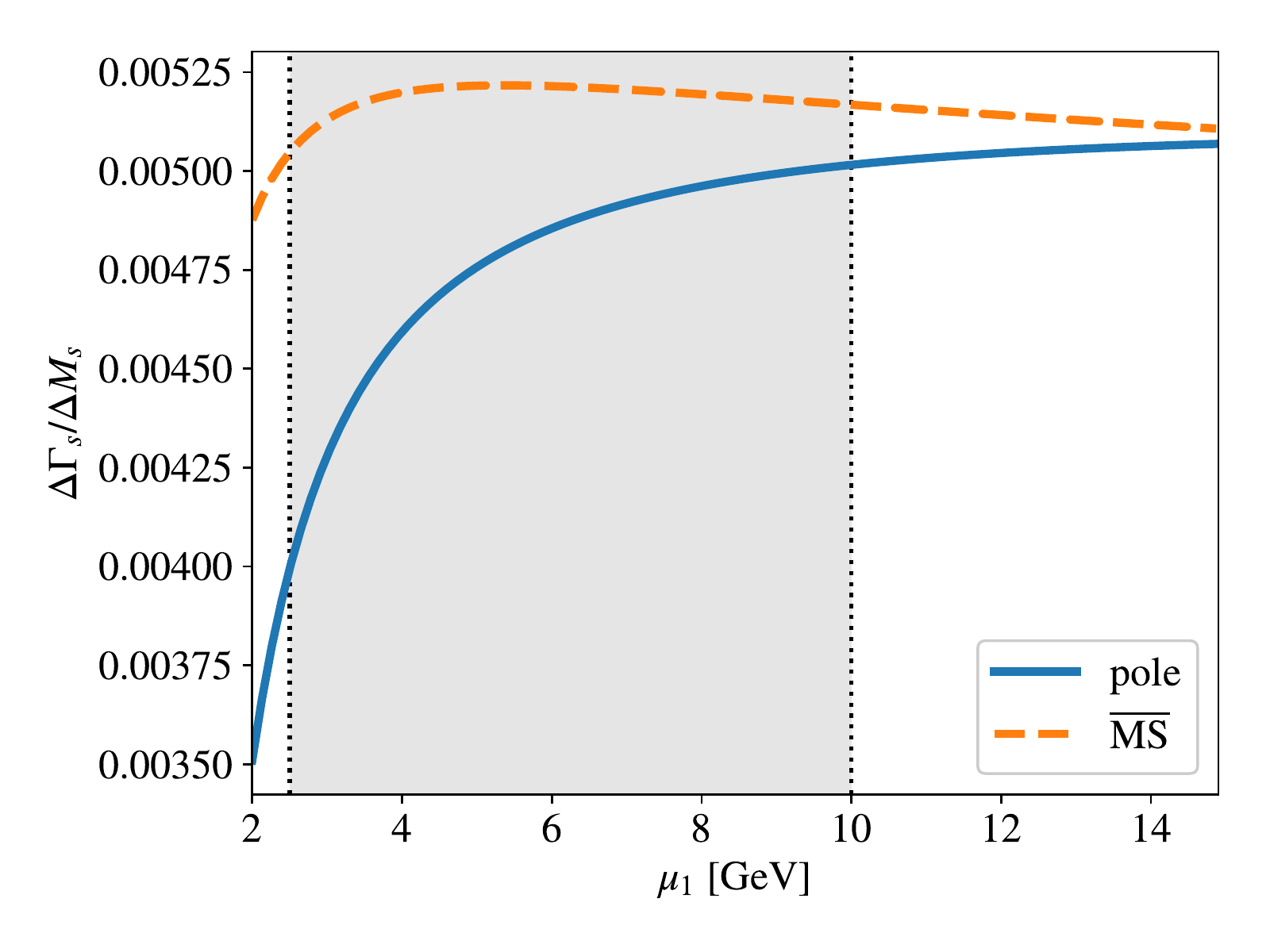}
      &
      \includegraphics[width=0.48\textwidth]{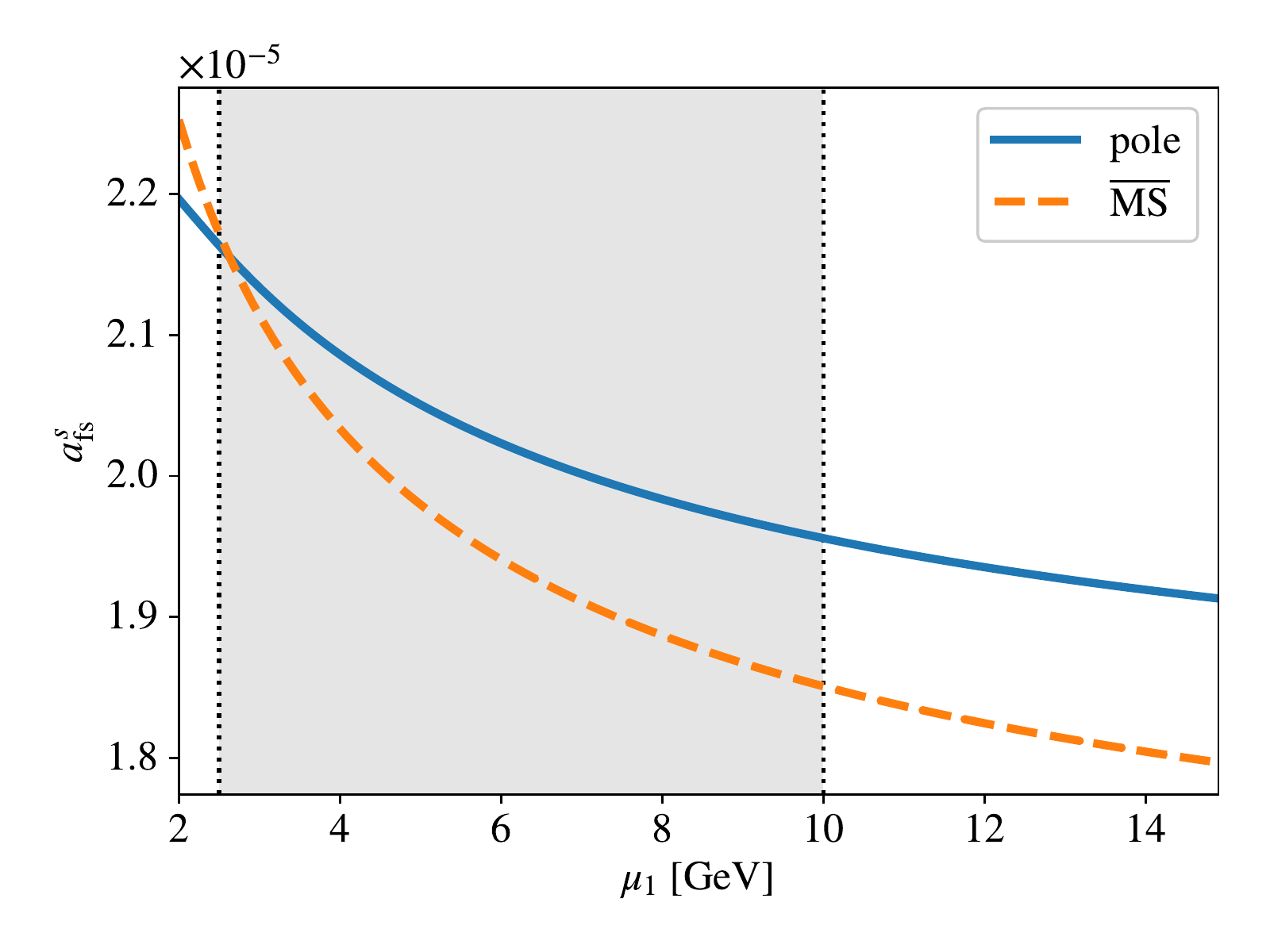}
        \\
      (a) & (b)
    \end{tabular}
  \end{center}
  \caption{\label{fig:scale} $\Delta\Gamma_s/\Delta M_s$ and
    $a^s_{\rm fs}$ as a function of $\mu_1$ for the $\overline{\rm MS}$
    (dashed orange) and pole (solid blue) renormalization schemes. The gray area
    shows the range of $\mu_1$ used to obtain the renormalization scale
    uncertainties quoted in \eqsand{eq::dGdM}{eq:afsnum}.} ~\\[-4mm] \hrule
\end{figure}  

The last three uncertainties in Eq.~(\ref{eq::dGdM}) are correlated
between the two schemes. The scale dependence is plotted in
\fig{fig:scale}(a) and leads to the asymmetric uncertainties quoted in
\eq{eq::dGdM}.  The difference between the central values found in the
pole and $\overline{\rm MS}$ schemes 
is around 11\%, {\it i.e.}\ of the expected size of an NNLO correction.

We proceed in a similar way for $a_{\rm fs}^s$. We
use Eq.~(\ref{eq::afsq}) and obtain
\begin{eqnarray}
  a_{\rm fs}^s
  &=&
      (2.07
      {}^{+0.10}_{-0.11}{}_{\rm scale}
      \pm 0.01_{B\tilde{B}_S}
      \pm 0.06_{1/m_b}
      \pm 0.06_{\textrm{input}}
  ) \times 10^{-5} \quad (\textrm{pole})
  \,,\nonumber\\ 
  a_{\rm fs}^s
  &=&
  (2.02
      {}^{{+0.15}}_{-0.17}{}_{\rm scale}               
      \pm 0.01_{B\tilde{B}_S} 
      \pm 0.05_{1/m_b}
      \pm 0.06_{\textrm{input}}
      ) \times 10^{-5}
  \quad (\overline{\textrm{MS}})
  \,. \label{eq:afsnum}
\end{eqnarray}
In \fig{fig:scale}(b) we show the dependence on $\mu_1$ for the two
renormalization schemes. Here the interval in the pole scheme is completely
contained in the one from the $\overline{\rm MS}$ scheme.

The predictions in \eqsand{eq::dGdM}{eq:afsnum} are consistent with
those of Ref.~\cite{Asatrian:2020zxa}, but the central values for
$\dg_s/\dm_s$ in \eq{eq::dGdM} are larger in both schemes. In
Ref.~\cite{Asatrian:2020zxa} only partial NLO corrections to the
$Q_{1,2}\times Q_{3-6}$ contribution and no $Q_{3-6}\times Q_{3-6}$ or
NNLO $Q_8$ terms have been included. Inspecting the sources of the
differences in detail, we find that almost 2/3 of these stem from the
new contributions presented in Ref.~\cite{Gerlach:2021xtb} and this
paper.  The remainder is due to terms, which are formally of higher order
in $\alpha_s$. Interestingly, the $\mu_1$ dependence of $\dg_s/\dm_s$ is
much smaller in \eq{eq::dGdM} compared to Ref.~\cite{Asatrian:2020zxa},
while the situation is vice versa for $ a_{\rm fs}^s$. We trace this
feature back to the use of $\alpha_s(\mu_1)$ versus $\alpha_s(\mu_2)$ in
certain NLO terms, both of which are allowed choices in the considered
order. In view of this observation and the fact that the intervals from the
scale uncertainty of $\dg_s/\dm_s$ in both schemes barely overlap,
we conclude that the $\mu_1$ dependence is not always a good estimate
of the size of the unknown higher-order corrections.

In a next step we can use the experimental result for $\Delta
M_s$~\cite{Zyla:2020zbs},
\begin{eqnarray}
  \Delta M_s^{\rm exp} &=& 17.7656 \pm 0.0057~\mbox{ps}^{-1}
                           \,,
\end{eqnarray}
and obtain for $\Delta\Gamma_s$ in the two renormalization schemes
\begin{eqnarray}
  \Delta\Gamma_s^{\rm pole} &=&   (0.083 ^{+0.005}_{-0.012}{}_{\rm scale} \pm
                                0.002_{B\tilde{B}_S} 
                            \pm 0.014_{1/m_b} \pm
                                0.001_{\textrm{input}})
                                ~\mbox{ps}^{-1}\,,\nonumber\\ 
  \Delta\Gamma_s^{\overline{\rm MS}} &=&   (0.092 ^{+0.0002}_{{-0.003}}{}_{\rm scale} \pm
                                         0.002_{B\tilde{B}_S} \pm
                                         0.012_{1/m_b} \pm
                                         0.001_{\textrm{input}})
                                         ~\mbox{ps}^{-1}\,. 
\end{eqnarray}

Comparing our prediction with the experimental value in \eq{eq:exp} we see
that both the ``pole'' and $\overline{\rm MS}$ results are consistent with the
measured value, but the central value of the former is closer to the
experimental result. One needs a better perturbative precision (which will
bring the ``pole'' and $\overline{\rm MS}$ results closer to each other and
reduce the scale uncertainty) and more precise lattice results for the matrix
elements of the $1/m_b$-suppressed operators to quantify new-physics
contributions to $\dg_s/\dm_s$.

The value for $\Delta\Gamma_s/\Delta M_s$ quoted in \eq{eq::dGdM} also applies
to $\Delta\Gamma_d/\Delta M_d$ for two reasons: First, while the
CKM-suppressed contribution to $\Delta\Gamma_d/\Delta M_d$ is a priori
expected to be relevant due to
$|\lambda^d_u/\lambda^d_t|\gg |\lambda^s_u/\lambda^s_t| $, it merely
contributes at the percent level because of a numerical cancellation in the
sum of $uc$ and $uu$ contributions \cite{Beneke:2003az}. Second, the
non-perturbative calculations of the $B_s$ and $B_d$ hadronic matrix elements
agree well within their error bars. As a result the central values for
$\Delta\Gamma_d/\Delta M_d$ and $\Delta\Gamma_s/\Delta M_s$ agree within a few
percent (see e.g. \cite{Asatrian:2020zxa}) and the difference is much smaller
than the uncertainty in \eq{eq::dGdM}. We find
\begin{eqnarray}
  \Delta\Gamma_d^{\rm pole} &\simeq& \frac{\Delta\Gamma_s}{\Delta
                                     M_s}\Big|_{\rm pole}
                                     \Delta M_d^{\rm exp} 
                                     \nonumber\\
                            &=&
                                (   {0.00238}
                                 ^{+0.00016}
                                 _{-0.00036}{}_{\rm scale}
                                \pm 0.00006_{B\tilde{B}_S} 
                                \pm 0.00040_{1/m_b} 
                                \pm 0.00003_{\textrm{input}})
                                ~\mbox{ps}^{-1}\,,\nonumber\\ 
  \Delta\Gamma_d^{\overline{\rm MS}} &\simeq& \frac{\Delta\Gamma_s}{\Delta
                                              M_s}\Big|_{\rm \overline{\rm MS}}
                                              \Delta M_d^{\rm exp} 
                                              \nonumber\\                                              
                            &=&
                                (   0.00264 
                                 ^{+0.00001}
                                 _{-0.00008}{}_{\rm scale}
                                \pm 0.00006_{B\tilde{B}_S} 
                                \pm 0.00034_{1/m_b} 
                                \pm 0.00003_{\textrm{input}})
                                ~\mbox{ps}^{-1}\,. 
\end{eqnarray}
where $\Delta M_d^{\rm exp}= (0.5065 \pm 0.0019)\, \mbox{ps}^{-1} $
\cite{hfag} has
been used.


\section{\label{sec::concl}Conclusions}

In this paper we have completed the calculation of the NLO contributions to
the decay matrix element $\Gamma_{12}^q$ appearing in \bbmq. These new
contributions involve two-loop diagrams with two four-quark penguin
operators. We have further calculated two-loop contributions with one or two
copies of the chromomagnetic penguin operators, which belong to NNLO or
N$^3$LO, respectively. 
All results are obtained as an expansion to first order in $z= m_c^2/m_b^2$,
except for the one-loop $Q_8 \times Q_8$ contribution for which our result has the
exact $z$-dependence.
With our new results the theoretical uncertainties
associated with the penguin sector are under full control and way below the
experimental error of the width difference $\dg_s\simeq 2|\Gamma_{12}^s|$ in
\eq{eq:exp}. We present updated predictions for $\dg_s$ and $\dg_d$ and the CP
asymmetry in flavor-specific $B_s$ decays, $a^s_{\rm fs}$. For the width
differences we find the predictions in the pole and $\overline{\rm MS}$
schemes to differ by 11\%, which invigorates the need for a full NNLO
calculation of the contributions from current-current operators.

We provide the newly obtained matching coefficients in a computer readable
format with full dependence on the number of colors $N_c$. In the same way we
present the renormalization matrix $Z_{ij}$ of the $\Delta B = 2$ operators
including the submatrices governing the mixing of evanescent operators with
physical operators and among each other.



\section*{Acknowledgements}  

We thank Artyom Hovhannisyan for useful discussions.
This research was supported by the Deutsche
Forschungsgemeinschaft (DFG, German Research Foundation) under grant 396021762
--- TRR 257 ``Particle Physics Phenomenology after the Higgs Discovery''.


\begin{appendix}


\section{\label{app::Z} Renormalization constants}

In this Appendix we describe the computation of the renormalization constants
required for the operator mixing in the $|\Delta B|=2$ theory and provide
explicit results relevant for the two-loop calculations presented in the main
part of this paper. Let us mention that all relevant renormalization constants
for the $|\Delta B|=1$ theory can be found in Ref.~\cite{Gambino:2003zm}.

For the computation of renormalization constants in the $\overline{\rm MS}$
scheme we can choose the external momenta and particle masses such, that the
amplitude $b + \bar{s} \to \bar{b} + s$ is infra-red finite.  This is possible
since $\overline{\rm MS}$ renormalization constants do not depend on kinematic
invariants and masses.  In our case it is convenient to set all
external momenta to zero and introduce a common mass for the strange and
bottom quark.  The gluon remains massless.  This leads to one-loop
vacuum integrals.

We work in a basis with physical operators $Q$, $\widetilde{Q}_S$
(c.f. Eq.~(\ref{eq::opDB2})) and $R_0$ and the corresponding evanescent
operators $E_1^{(1)}, \ldots, E_5^{(1)}$ from Eq.~(\ref{eq:E1}).  We have to
introduce further evanescent operators, which contains the Dirac structures
present in the $\Delta B=1$ amplitude.  As can be seen from
Tab.~\ref{tab::gamma_structures} the contribution $Q_{3-6}\times Q_{3-6}$ has
the largest number of $\gamma$ matrices and requires that the evanescent operators
$E_i^{(4)}$ (see Eq.~(\ref{eq::evOp_add})) are taken into account in the computation of
the amplitude. The same evanescent operators are also needed for the
computation of the renormalization constants. In analogy to the amplitude
calculation, also for the renormalization constants the ${\cal O}(\epsilon)$
terms $e_{i,j}$ defined in Eq.~(\ref{eq::evOp_add}) are only needed for $E_i^{(1)}$.

We can write the matrix of renormalization constants as a $20\times 20$
matrix which is naturally decomposed into four sub-matrices
\begin{eqnarray}
  Z_{\Delta B=2} &=&
                     \left(
                     \begin{array}{cc} 
                       Z_{QQ} & Z_{QE} \\
                       Z_{EQ} & Z_{EE}
                     \end{array}
                                \right)\,,
\end{eqnarray}
where $Z_{QQ}$, $Z_{QE}$, $Z_{EQ}$ and $Z_{EE}$ have the dimension
$3\times 3$, $3\times 17$, $17\times 3$ and $17\times 17$, respectively. We
define $Z_{\Delta B=2}$ via the renormalization of
the coefficient functions as follows
\begin{eqnarray}
  \vec{C}^{\rm bare} &=& Z_{\Delta B=2}^T \vec{C}^{\rm ren}\,,
\end{eqnarray}
where $\vec{C}^{\rm bare}$ and $\vec{C}^{\rm ren}$ are 20-dimensional vectors
of the bare and renormalized $|\Delta B|=2$ coefficient functions,
respectively.  The perturbative expansion of the sub-matrices is introduced as
\begin{eqnarray}
  Z_{QQ} &=& 1 +  \frac{\alpha_s}{4\pi} \frac{1}{\epsilon}
             Z_{QQ}^{(1,1)}\,,
             \nonumber\\
  Z_{QE} &=&  \frac{\alpha_s}{4\pi} \frac{1}{\epsilon} Z_{QE}^{(1,1)}\,,
             \nonumber\\
  Z_{EE} &=& 1+   \frac{\alpha_s}{4\pi} \frac{1}{\epsilon}
             Z_{EE}^{(1,1)}\,, 
             \nonumber\\
  Z_{EQ} &=&  \frac{\alpha_s}{4\pi} Z_{EQ}^{(1,0)}\,,
\end{eqnarray}
where the first superscript denotes the order in $\alpha_s$ and the second one
the order in $1/\epsilon$. Note that at one-loop order the matrix
$Z_{EQ}$ only contains finite contributions.

In order to determine the matrix elements of $Z_{\Delta B=2}$ we compute
the amplitude $b + \bar{s} \to \bar{b} + s$ in the kinematics described
above, take into account the field renormalization of the external
quarks in the $\overline{\rm MS}$ scheme and require that the remaining
poles in $\epsilon$, which are all of ultra-violet nature, are absorbed
by the operator mixing via $Z_{\Delta B=2}$.  This condition fixes all
matrix elements but the ones in $Z_{EQ}$. The latter are fixed by the
requirement that the contributions of evanescent operators vanish in
$D=4$ dimensions~\cite{Buras:1989xd,Herrlich:1994kh}. Note that to our
order we do not have to renormalize the common strange and bottom quark
mass.

An important check of our calculation is the locality of the
extracted renormalization constants. Furthermore, we perform the
calculation for general QCD gauge parameter and observe that
the matrix  $Z_{\Delta B=2}$  is independent of $\xi$.

In the following we present explicit results for the one-loop
corrections to $Z_{QQ}$, $Z_{QE}$ and $Z_{EE}$.
For $N_c=3$ we have
\begin{align}
  Z_{QQ}^{(1,1)} &=  \left(
                   \begin{array}{ccc}
                     2 & 0 & 0 \\
                     -\frac{4}{3} & \frac{8}{3} & \frac{8}{3} \\
                     2 & 8 & -2 \\
                   \end{array}
  \right) \,,
  \\
  Z_{QE}^{(1,1)} &=  \left(
                   \begin{array}{ccccccccccccccccc}
                     3 & \frac{1}{2} & -\frac{1}{6} & 0 & 0 & 0 & 0 & 0 & 0 & 0 & 0 & 0 & 0 & 0 & 0 & 0 & 0 \\
                     0 & 0 & 0 & -\frac{7}{12} & -\frac{1}{4} & 0 & 0 & 0 & 0 & 0 & 0 & 0 & 0 & 0 & 0 & 0 & 0 \\
                     \frac{3}{2} & \frac{1}{4} & -\frac{1}{12} & -\frac{13}{12} & -\frac{1}{12} & 0 & 0 & 0 & 0 & 0 & 0 & 0 & 0 & 0 & 0 & 0 & 0 \\
                   \end{array}
  \right)\,,
\end{align}

{\scalefont{1.0}
\begin{equation}
  \rotatebox{90}{$
  Z_{EE}^{(1,1)} =  \left(
                   \begin{array}{ccccccccccccccccc}
                     -4 & \frac{1}{12} & \frac{5}{12} & 0 & 0 & 0 & 0 & 0 & 0 & 0 & 0 & 0 & 0 & 0 & 0 & 0 & 0 \\
                     0 & -\frac{59}{3} & -5 & 0 & 0 & \frac{7}{12} & \frac{1}{4} & 0 & 0 & 0 & 0 & 0 & 0 & 0 & 0 & 0 & 0 \\
                     0 & -13 & \frac{13}{3} & 0 & 0 & \frac{1}{2} & -\frac{1}{6} & 0 & 0 & 0 & 0 & 0 & 0 & 0 & 0 & 0 & 0 \\
                     0 & 0 & 0 & -22 & -\frac{2}{3} & 0 & 0 & -\frac{1}{4} & -\frac{7}{12} & 0 & 0 & 0 & 0 & 0 & 0 & 0 & 0 \\
                     0 & 0 & 0 & -\frac{44}{3} & 4 & 0 & 0 & \frac{1}{6} & -\frac{1}{2} & 0 & 0 & 0 & 0 & 0 & 0 & 0 & 0 \\
                     0 & -\frac{1888}{3} & 96 & 0 & 0 & \frac{41}{3} & -9 & 0 & 0 & \frac{7}{12} & \frac{1}{4} & 0 & 0 & 0 & 0 & 0 & 0 \\
                     0 & -288 & \frac{1568}{3} & 0 & 0 & 3 & -\frac{67}{3} & 0 & 0 & \frac{1}{2} & -\frac{1}{6} & 0 & 0 & 0 & 0 & 0 & 0 \\
                     0 & 0 & 0 & \frac{608}{3} & -\frac{544}{3} & 0 & 0 & -\frac{22}{3} & -2 & 0 & 0 & -\frac{1}{6} & \frac{1}{2} & 0 & 0 & 0 & 0 \\
                     0 & 0 & 0 & \frac{992}{3} & -\frac{160}{3} & 0 & 0 & -4 & -\frac{4}{3} & 0 & 0 & \frac{1}{4} & \frac{7}{12} & 0 & 0 & 0 & 0 \\
                     0 & -\frac{39424}{3} & 2560 & 0 & 0 & -672 & 224 & 0 & 0 & \frac{185}{3} & -25 & 0 & 0 & \frac{7}{12} & \frac{1}{4} & 0 & 0 \\
                     0 & -5632 & \frac{34304}{3} & 0 & 0 & -224 & 672 & 0 & 0 & 19 & -\frac{211}{3} & 0 & 0 & \frac{1}{2} & -\frac{1}{6} & 0 & 0 \\
                     0 & 0 & 0 & \frac{7424}{3} & -\frac{1792}{3} & 0 & 0 & 720 & -240 & 0 & 0 & -\frac{130}{3} & 10 & 0 & 0 & -\frac{1}{6} & \frac{1}{2} \\
                     0 & 0 & 0 & \frac{8960}{3} & -\frac{256}{3} & 0 & 0 & 240 & -720 & 0 & 0 & -16 & \frac{104}{3} & 0 & 0 & \frac{1}{4} & \frac{7}{12} \\
	\ast & \ast & \ast & \ast & \ast & \ast & \ast & \ast & \ast & \ast & \ast & \ast & \ast & \ast & \ast & \ast & \ast \\
	\ast & \ast & \ast & \ast & \ast & \ast & \ast & \ast & \ast & \ast & \ast & \ast & \ast & \ast & \ast & \ast & \ast \\
	\ast & \ast & \ast & \ast & \ast & \ast & \ast & \ast & \ast & \ast & \ast & \ast & \ast & \ast & \ast & \ast & \ast \\
	\ast & \ast & \ast & \ast & \ast & \ast & \ast & \ast & \ast & \ast & \ast & \ast & \ast & \ast & \ast & \ast & \ast \\
\end{array}
\right)
$}
\end{equation}
}
where the entries ``$\ast$'' are not needed for our calculation.

The (finite) matrix $Z_{EQ}^{(1,0)}$ depends on the
${\cal O}(\epsilon)$ terms of the evanescent operators, $e^{(i)}_{j}$ and
$e^{(i)}_{j,k}$. It is given by
\begin{eqnarray}
	Z_{EQ}^{(1,0)} &=& \begin{pmatrix}
		Z_{EQ,1}^{(1,0)} & Z_{EQ,2}^{(1,0)} & Z_{EQ,3}^{(1,0)}
	\end{pmatrix}\,,
\end{eqnarray}
where
{
	\footnotesize
\begin{align}
		Z_{EQ,1}^{(1,0)} &=
\left(
\begin{array}{c}
	0 \\
	\frac{7}{12} e_1^{ (2)}+\frac{1}{4} e_2^{ (2)}+\frac{464}{3} \\
	\frac{1}{2} e_1^{ (2)}-\frac{1}{6} e_2^{ (2)}-\frac{16}{3} \\
	\frac{1}{8} e_{3,2}^{ (2)}+\frac{7}{24} e_{4,2}^{ (2)}+84 \\
	-\frac{1}{12} e_{3,2}^{ (2)}+\frac{1}{4} e_{4,2}^{ (2)}+44 \\
	\frac{35}{3} e_1^{ (2)}+\frac{7}{12} e_1^{ (3)}-9 e_2^{ \
		(2)}+\frac{1}{4} e_2^{ (3)}+\frac{17920}{3} \\
	3 e_1^{ (2)}+\frac{1}{2} e_1^{ (3)}-\frac{73}{3} e_2^{ \
		(2)}-\frac{1}{6} e_2^{ (3)}-\frac{14336}{3} \\
	\frac{4}{3} e_{3,1}^{ (2)}+\frac{4}{3} e_{3,2}^{ (2)}+e_{4,2}^{ \
		(2)}+\frac{1}{12} e_{3,2}^{ (3)}-\frac{1}{4} e_{4,2}^{ (3)}+64 \\
	2 e_{3,2}^{ (2)}-\frac{1}{8} e_{3,2}^{ (3)}+\frac{4}{3} e_{4,1}^{ \
		(2)}-\frac{5}{3} e_{4,2}^{ (2)}-\frac{7}{24} e_{4,2}^{ (3)}-1728 \\
	-672 e_1^{ (2)}+\frac{179}{3} e_1^{ (3)}+\frac{7}{12} e_1^{ (4)}+224 \
	e_2^{ (2)}-25 e_2^{ (3)}+\frac{1}{4} e_2^{ (4)}+\frac{901120}{3} \\
	-224 e_1^{ (2)}+19 e_1^{ (3)}+\frac{1}{2} e_1^{ (4)}+672 e_2^{ (2)}-\
	\frac{217}{3} e_2^{ (3)}-\frac{1}{6} e_2^{ (4)}-\frac{843776}{3} \\
	-360 e_{3,2}^{ (2)}+\frac{58}{3} e_{3,2}^{ (3)}+\frac{1}{12} \
	e_{3,2}^{ (4)}+120 e_{4,2}^{ (2)}+\frac{4}{3} e_{3,1}^{ (3)}-5 \
	e_{4,2}^{ (3)}-\frac{1}{4} e_{4,2}^{ (4)}+24064 \\
	-120 e_{3,2}^{ (2)}+8 e_{3,2}^{ (3)}-\frac{1}{8} e_{3,2}^{ (4)}+360 \
	e_{4,2}^{ (2)}+\frac{4}{3} e_{4,1}^{ (3)}-\frac{59}{3} e_{4,2}^{ \
		(3)}-\frac{7}{24} e_{4,2}^{ (4)}-44544 \\
	\ast \\
	\ast \\
	\ast \\
	\ast \\
\end{array}
\right),
\end{align}
}

{
\scalefont{0.5}
\begin{align}
	Z_{EQ,2}^{(1,0)} &=
	\left(
	\begin{array}{c}
		0 \\
		0 \\
		0 \\
		-\frac{1}{4} e_{3,1}^{ (2)}+\frac{1}{4} e_{3,2}^{ (2)}-\frac{7}{12} e_{4,1}^{ (2)}+\frac{7}{12} e_{4,2}^{ (2)}+32 \\
		\frac{1}{6} e_{3,1}^{ (2)}-\frac{1}{6} e_{3,2}^{ (2)}-\frac{1}{2} e_{4,1}^{ (2)}+\frac{1}{2} e_{4,2}^{ (2)}+32 \\
		0 \\
		0 \\
		-10 e_{3,1}^{ (2)}-\frac{1}{6} e_{3,1}^{ (3)}+2 e_{3,2}^{ (2)}-2 e_{4,1}^{ (2)}+2 e_{4,2}^{ (2)}+\frac{1}{6} e_{3,2}^{ (3)}+\frac{1}{2} e_{4,1}^{ (3)}-\frac{1}{2} e_{4,2}^{ (3)}-1536 \\
		-4 e_{3,1}^{ (2)}+\frac{1}{4} e_{3,1}^{ (3)}+4 e_{3,2}^{ (2)}-4 e_{4,1}^{ (2)}-4 e_{4,2}^{ (2)}-\frac{1}{4} e_{3,2}^{ (3)}+\frac{7}{12} e_{4,1}^{ (3)}-\frac{7}{12} e_{4,2}^{ (3)}-1536 \\
		0 \\
		0 \\
		720 e_{3,1}^{ (2)}-46 e_{3,1}^{ (3)}-\frac{1}{6} e_{3,1}^{ (4)}-720 e_{3,2}^{ (2)}-240 e_{4,1}^{ (2)}+240 e_{4,2}^{ (2)}+38 e_{3,2}^{ (3)}+10 e_{4,1}^{ (3)}-10 e_{4,2}^{ (3)}+\frac{1}{6} e_{3,2}^{ (4)}+\frac{1}{2} e_{4,1}^{ (4)}-\frac{1}{2} e_{4,2}^{ (4)}-12288 \\
		240 e_{3,1}^{ (2)}-16 e_{3,1}^{ (3)}+\frac{1}{4} e_{3,1}^{ (4)}-240 e_{3,2}^{ (2)}-720 e_{4,1}^{ (2)}+720 e_{4,2}^{ (2)}+16 e_{3,2}^{ (3)}+32 e_{4,1}^{ (3)}-40 e_{4,2}^{ (3)}-\frac{1}{4} e_{3,2}^{ (4)}+\frac{7}{12} e_{4,1}^{ (4)}-\frac{7}{12} e_{4,2}^{ (4)}-12288 \\
	\ast \\
	\ast \\
	\ast \\
	\ast \\
	\end{array}
	\right),
\end{align}
}

{
	\footnotesize
\begin{align}
        Z_{EQ,3}^{(1,0)}
	&=
	\left(
	\begin{array}{c}
		0 \\
		0 \\
		0 \\
		-\frac{1}{4} e_{3,2}^{ (2)}-\frac{7}{12} e_{4,2}^{ (2)}-168 \\
		\frac{1}{6} e_{3,2}^{ (2)}-\frac{1}{2} e_{4,2}^{ (2)}-88 \\
		0 \\
		0 \\
		-\frac{8}{3} e_{3,1}^{ (2)}-\frac{8}{3} e_{3,2}^{ (2)}-2 e_{4,2}^{ (2)}-\frac{1}{6} e_{3,2}^{ (3)}+\frac{1}{2} e_{4,2}^{ (3)}-128 \\
		-4 e_{3,2}^{ (2)}+\frac{1}{4} e_{3,2}^{ (3)}-\frac{8}{3} e_{4,1}^{ (2)}+\frac{10}{3} e_{4,2}^{ (2)}+\frac{7}{12} e_{4,2}^{ (3)}+3456 \\
		0 \\
		0 \\
		720 e_{3,2}^{ (2)}-\frac{116}{3} e_{3,2}^{ (3)}-\frac{1}{6} e_{3,2}^{ (4)}-240 e_{4,2}^{ (2)}-\frac{8}{3} e_{3,1}^{ (3)}+10 e_{4,2}^{ (3)}+\frac{1}{2} e_{4,2}^{ (4)}-48128 \\
		240 e_{3,2}^{ (2)}-16 e_{3,2}^{ (3)}+\frac{1}{4} e_{3,2}^{ (4)}-720 e_{4,2}^{ (2)}-\frac{8}{3} e_{4,1}^{ (3)}+\frac{118}{3} e_{4,2}^{ (3)}+\frac{7}{12} e_{4,2}^{ (4)}+89088 \\
	\ast \\
	\ast \\
	\ast \\
	\ast \\
	\end{array}
	\right).
\end{align}
}


\end{appendix}



\end{document}